\documentclass[12pt,preprint]{aastex}

\shorttitle{Palomar 13 Binaries and Blue Straggler Stars}
\shortauthors{}

\begin{document}

\title{The Blue Straggler and Main-sequence Binary Population of the Low-Mass Globular 
Cluster Palomar 13\footnote{Based on data obtained at the W.M. Keck
Observatory, which is operated as a scientific partnership among the
University of California, the California Institute of Technology, and
NASA, and was made possible through the generous financial support of
the W.M. Keck Foundation}}

\author{L. Lee Clark,  Eric L. Sandquist}
\affil{Astronomy Department, San Diego State University, San Diego, CA 92182}
\email{clark@mintaka.sdsu.edu, erics@mintaka.sdsu.edu}

\author{Michael Bolte} 
\affil{University of California Observatories, Lick Observatory, Board
of Studies in Astronomy and Astrophysics, University of California,
Santa Cruz, CA 95064}
\email{bolte@ucolick.org}

\begin{abstract}
We present high-precision $VI$ photometry of stars from the middle of
the giant branch to about 5 magnitudes below the main-sequence turnoff
in the globular cluster Palomar 13 based on images obtained with the
Keck II 10m telescope. We tabulate a complete sample of blue
stragglers in the cluster out to about 18 core radii.  The blue
straggler population is significantly more centrally concentrated than
the giant star sample, which is in turn significantly more centrally
concentrated than the main-sequence star sample. Palomar 13 has one of
the highest specific frequencies of blue stragglers of any known
globular cluster, but the specific frequency of blue stragglers in the
outskirts of the cluster does not increase as has been seen in denser
clusters.  We also identify a group of faint blue stragglers (bluer
than the turnoff, but having about the same magnitude) that outnumbers
the brighter stragglers by more than a factor of 2.  The cluster's
color-magnitude diagram shows a large excess of stars to the red of
the main sequence, indicating that the cluster's binary fraction is at
least $(30\pm4)$\%, which appears to be similar to that of the
low-mass cluster E3 but significantly higher than that of the more
massive clusters Pal 5 and NGC 288.
\end{abstract}


\section{Introduction} 

The low-density globular cluster Palomar 13 was discovered on a
National Geographic Society-Palomar Observatory Sky Survey
photographic plate by Wilson (1955) while searching for Sculptor-type
dwarf galaxies within the Local Group. Early color-magnitude studies 
showed it to be at large galactocentric distance ($\sim 25$ kpc) and very
low luminosity with an integrated absolute magnitude $M_V \sim -3.4$ 
(Ciatti, Rosino, \& Sussi 1965; Ortolani, Rosino, \& Sandage 1985) 

Recent studies of Palomar 13 have found indications that the cluster is
probably in the late stages of tidal destruction by the Galaxy. Siegel
et al. (2001; hereafter SMCT) was the first to measure proper motions
of Palomar 13 stars in order to estimate membership
probabilities. They found evidence of member stars beyond the
cluster's limiting radius, possibly the result of evaporation or tidal
stripping. \citet{cot02} presented photometry and radial velocity
measurements supporting the idea that the cluster is either in the
process of being tidally disrupted, or that its dynamics are dominated
by a dark matter halo. Both studies noted a ``secondary sequence''
beside the single-star fiducial line in the color-magnitude diagram,
probably indicating a substantial binary population within the
cluster. The existence of such a binary population is consistent with
the idea that dynamical processes have led to mass segregation with 
the preferential loss of 
low-mass single stars and the retention of more massive binary star
systems in the cluster core. However, neither of these studies
attempted to characterize the size of the binary population.

Several previous studies \citep{bor97,sie01,cot02} have noted the
presence of blue straggler stars --- in fact, the blue stragglers
outnumber the horizontal branch stars in the cluster. It has been
recognized for some time that the lower-mass clusters are particularly
efficient at producing blue stragglers \citep{fer95,pio04} 
To date though, {\it complete} surveys of blue stragglers
have been most common in more massive clusters like M3 \citep{fer93} and 
47 Tuc \citep{fer04}. 
The leading explanations for the creation of blue stragglers involve
stellar collisions or primordial binary coalescence. Because the dynamics of
a cluster will affect the number, period distribution, and evolution
of its binary stars, 
a dynamically mature low-mass, low-density cluster is
likely to have a different binary population than will be observed in more massive clusters.

In this paper, we present the results of our photometric study of
Palomar 13 examining the effects that the cluster's dynamics may have
had on the stellar population.  The greatest obstacle is Palomar 13's
distance. But Palomar 13 provides an excellent opportunity to study
blue stragglers: it has little reddening and low field-star
contamination (thanks to its high galactic latitude), proper motion
membership information is available for stars at the main sequence
turnoff and brighter (SMCT), and the low stellar density makes
accurate photometry easy over the cluster's entire radial extent. On
the other hand, an examination of the binary content of the cluster
has a few difficulties.  While spectra have been obtained for some giant
stars \citet{cot02}, large surveys for radial velocity variations are
impractical due to the cluster's distance.  Photometric selection of
probable binary stars is more promising, but care must be taken to
minimize the small (but nonzero) field star population. Membership
information is available from radial velocities and proper motions,
but only for stars brighter than about the main sequence turnoff. (In
addition, the radial velocity of Pal 13 is quite low, so that there is
a relatively small difference between the radial velocities of cluster
and field stars \citep{cot02}.) To understand the main sequence binary
population, we will need to rely on high-quality photometry, both to
identify probable binaries and to eliminate as many field stars as
possible.

Section \ref{2:orc} gives a brief description of our observations and
the data reduction procedures. Section \ref{3:one} gives a brief discussion
of the color-magnitude diagram (CMD). In section \ref{4:bss}, we
present a thorough study of the blue straggler population within
Palomar 13 and compare with other Milky Way globular clusters. In section
\ref{binary} we place limits on the binary content of the cluster.

\section{Observations and Data Reduction}\label{2:orc} 

We used the Echellette Spectrograph and Imager (ESI, Sheinis et al. 2002) 
in direct imaging
mode on the night of 1999 September 2 to observe Palomar 13 at the 10m
Keck II telescope at the Cassegrain focus. ESI in its direct
imager mode has a $2\arcmin \times 4\arcmin$ field with a
$0\farcs15$ pix$^{-1}$ scale. The seeing during the run was
excellent, with the FWHM of stars ranging between about $0\farcs35$ and
$0\farcs7$. We collected one 100 s and three 600 s exposures in $V$, and one
60 s and three 600 s exposures in $I$. In comparison, \citet{cot02} only had two
exposures in each filter (10 and 300 s in $V$, and 6 and 180 s in $I$). The
seeing for their frames ranged from 0\farcs7 to 0\farcs8. Each of these factors
contributed to a noticeable improvement in the photometric scatter on the lower main sequence
in our photometry compared to theirs.

The Keck cluster images were bias-subtracted and flat-fielded using
IRAF\footnote{IRAF (Image Reduction and Analysis Facility) is
distributed by the National Optical Astronomy Observatories, which are
operated by the Association of Universities for Research in Astronomy,
Inc., under contract with the National Science Foundation.}. The
photometric measurements were made using point-spread function fitting
in the programs DAOPHOT II and ALLSTAR (Stetson 1993). All of the
frames were reduced once, and object-subtracted frames from ALLSTAR
were examined to recover additional stars. A new master list of stars
was then derived from those objects that were found in all three 600 s
$I$-band exposures. This list was then merged with the list of objects
found in all three 600 s $V$-band exposures (which did not go as faint
as the $I$-band frames), and the combined list was used in final runs
of ALLSTAR on the long exposure frames. The final star list was made
by finding stars that were detected in both filter bands.

We calibrated our
photometry to that of \citet{cot02}, whose dataset had a comparable
faint limit to ours, and who were able to observe Landolt (1992) 
standard fields interspersed with their cluster field.
We chose stars with $V < 23$
and $(V-I) < 1.05$ for the calibration.
Although our absolute photometry is not 
independent of theirs, our {\it relative} photometry
is an improvement over that of \citet{cot02}. 

\begin{figure}
\plotone{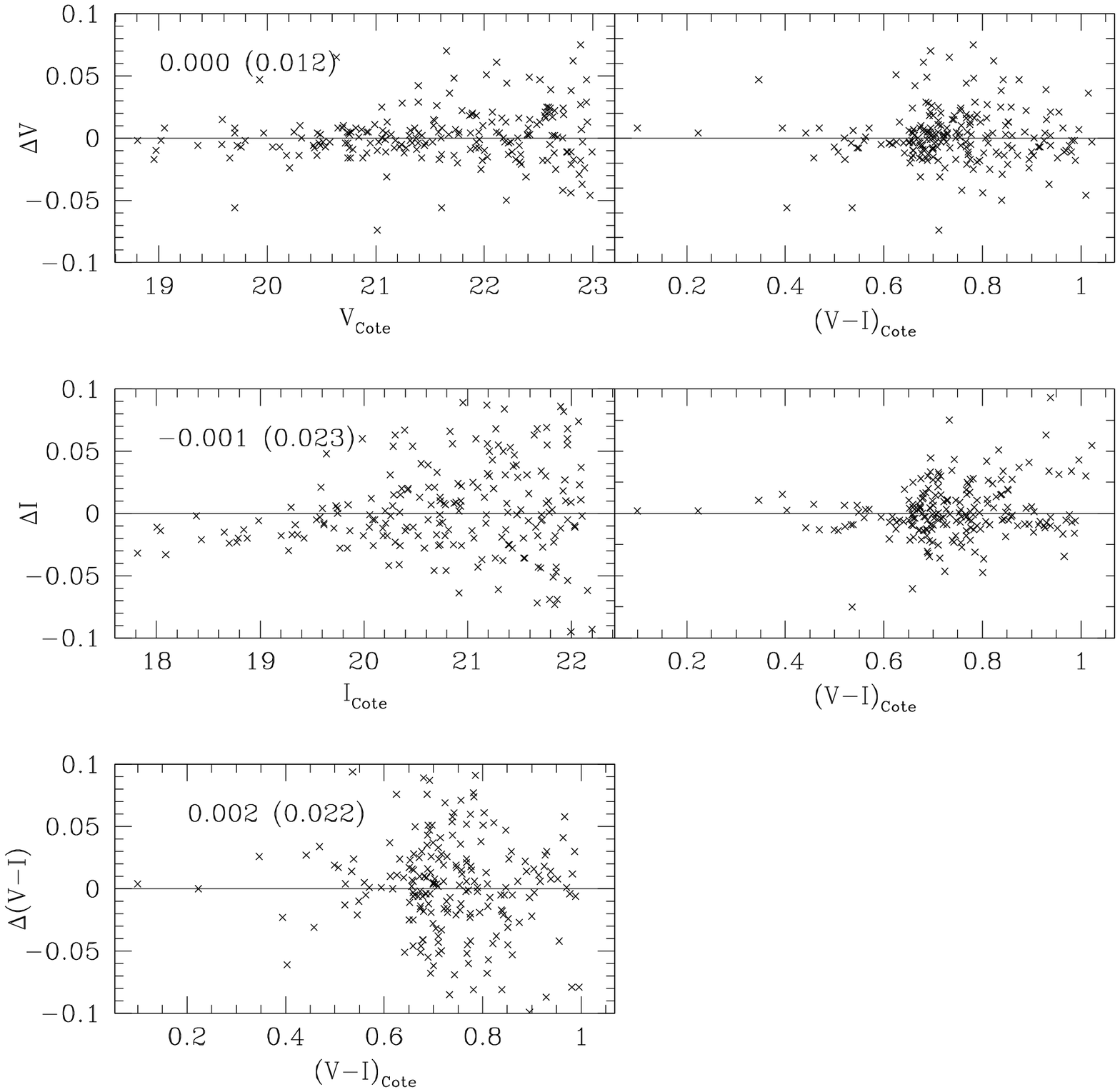}
\caption{Residuals for $V$ and $I$ measurements of the 211 Palomar 13
stars used in our calibration (in the sense of our magnitudes minus
\citet{cot02} magnitudes). The median residuals and semi-interquartile
range (a measure of scatter) are plotted in the panels on the
left. \label{fig2}}
\end{figure}

The equations used in the photometric transformations are:
\begin{equation} \label{e1}
v=V+a_{0}+a_{1} (V-I)+a_{2} (V-I)^2
\end{equation}
\begin{equation} \label{e2}
i=I+b_{0}+b_{1} (V-I)
\end{equation} 
where $v$ and $i$ are instrumental magnitudes, and $V$ and $I$ are
standard magnitudes. The color transformation coefficients were $a_1 =
0.100 \pm 0.040$, $a_2 = -0.037 \pm 0.029$ and $b_1 = 0.107 \pm 0.015$. 
A comparison of our calibrated
photometry with that of \citet{cot02}
is shown in Fig. \ref{fig2}, along with median residuals. Our photometry is very
accurately placed on the same system as that of \citet{cot02}, with
the exception of a slight ($\sim 0.02 - 0.03$ mag) systematic trend in
$I$ for stars on the giant branch. \citet{cot02} compared their
photometry with that of previous studies, finding good agreement with
that of \citet{sie01}, but finding significant offsets and trends with
magnitude and color compared to earlier photometric studies
\citep{ort85,bor97}.

\section{The Color-Magnitude Diagram (CMD)} \label{3:one}

\begin{figure}
\plotone{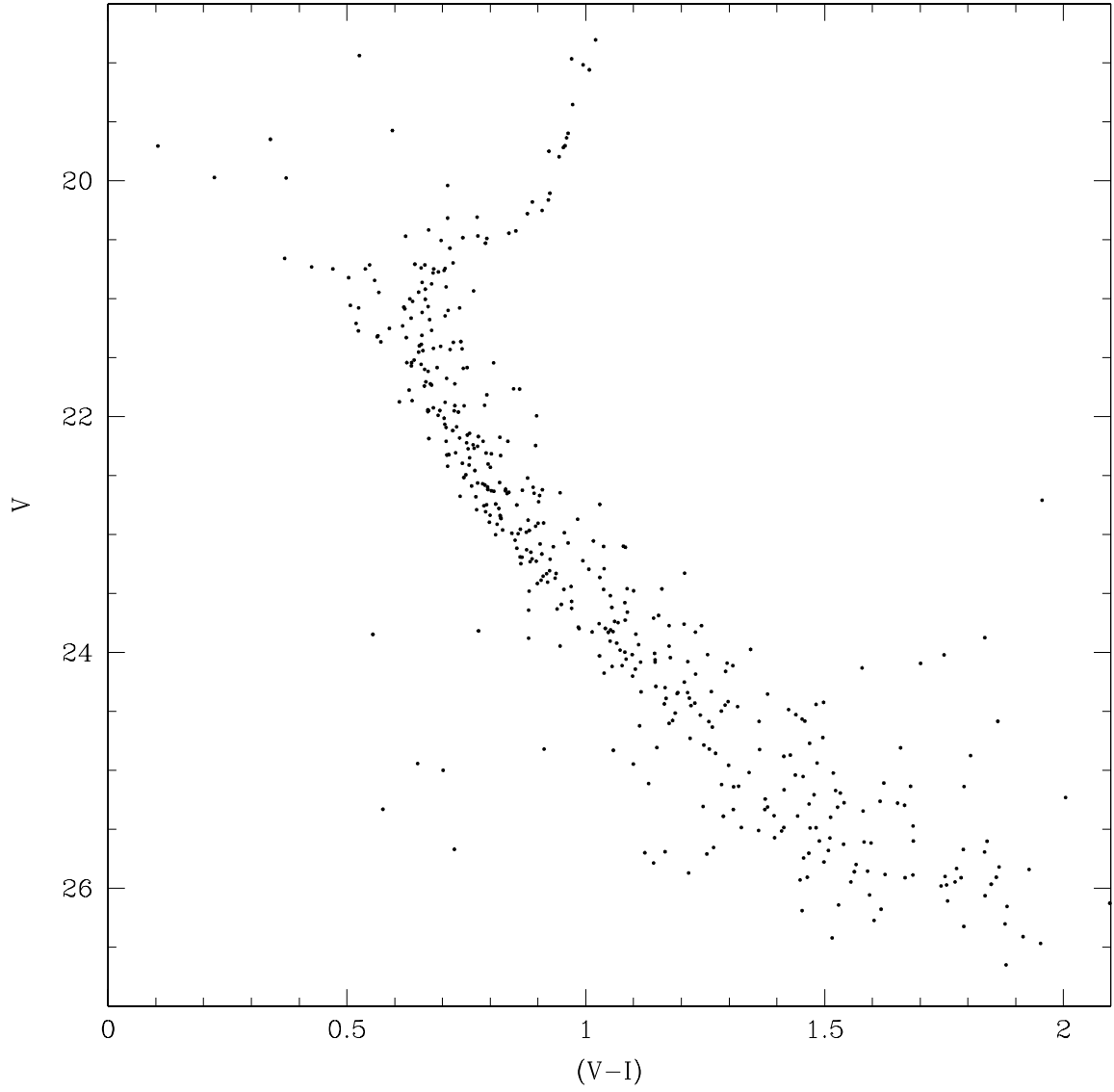}
\caption{The color-magnitude diagram for Palomar 13 from our Keck 
photometry.\label{fig6}}
\end{figure}

\begin{figure}
\plotone{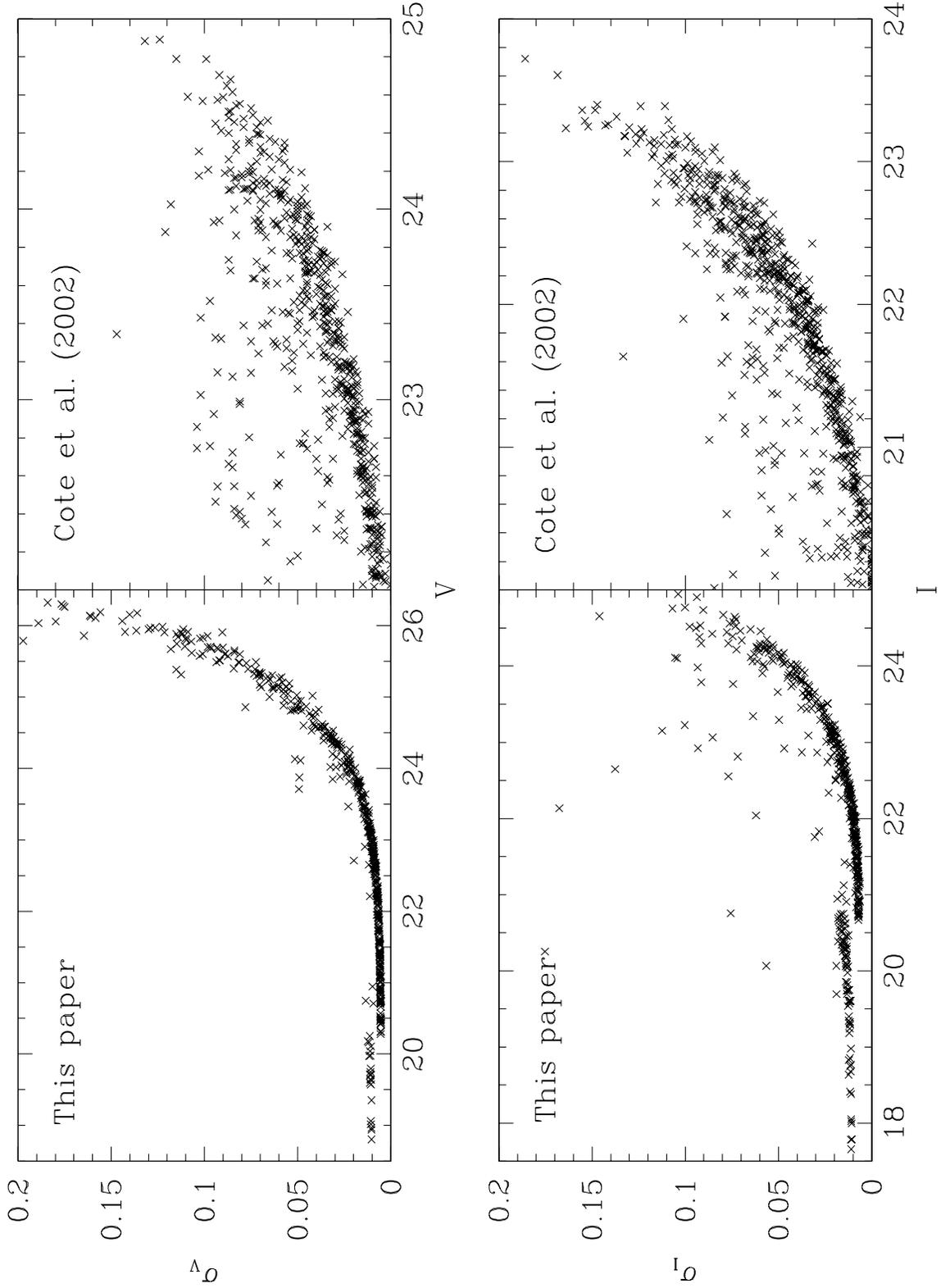}
\caption{Formal measurement errors for the photometry from this study
({\it left panels}) and \citet{cot02} ({\it right
panels}). The jumps seen at $V \approx 20.25$ and $I \approx 20.6$
mark the saturation limits for the long exposures in this
study.\label{errs}}
\end{figure}

Fig. \ref{fig6} shows the CMD of the 478 stars detected in both
filters with $18.5 \la V \la 26.5$. The main source of contamination
in the CMD for this cluster was background galaxies. We therefore
employed a cut on the image sharpness provided by ALLSTAR for each
object: $-0.5 < \mbox{SHARP} < 0.5$. This cut only affected objects
fainter than the cluster turnoff ($V_{TO} \approx 21.15 \pm
0.10$). Bright cluster stars were saturated in all of our Keck
images. The photometric precision for main-sequence stars is
relatively high, as can be seen by comparing the lower main sequence
in Fig. \ref{fig6} and Fig. 4 of \citet{cot02}. In Fig. \ref{errs}, we
present plots comparing formal measurement errors in $V$ and $I$ for
the measurements from this studay and that of \citet{cot02}. The
reduction in the photometric scatter allows us to clearly see a
substantial population of stars to the red of the single-star main
sequence. This group of stars is interpreted as a binary star
population as will be discussed further in section \ref{binary}.

Our photometry sample is smaller than that of \citet{cot02} due to
smaller field size. Because \citet{cot02} discussed the main-sequence
luminosity function for Palomar 13, we will not present a new analysis
here. Regardless, the luminosity function is
nearly flat with magnitude, a fact which can almost be seen by
eye in the CMD.  This is cited as evidence of the effectiveness of
mass segregation and the evaporation of
low-mass stars. Similar flat luminosity functions have been measured for
other more massive clusters like Palomar 5 \citep{gri01,oden} and NGC 6752 \cite{rub99}.


\section{Straggler Stars and Bright Binary Systems} \label{4:bss}

\subsection{Background}

The leading explanations for blue straggler star formation involve the
creation of a star more massive than the turnoff mass after the epoch
of star formation in the cluster. Primordial binaries evolving into
contact are the leading candidates for forming field blue stragglers
--- spectroscopic binaries with low eccentricity found among halo blue
stragglers imply that mass transfer is important (Carney et
al. 2001). In the denser environments of globular clusters, binary
star evolution may still play an important role, but stellar
collisions (involving single or binary stars) are also likely to form
some fraction of the population.  Efforts are now being made to
identify the relative roles of these different formation
mechanisms. For example, the specific frequency of blue stragglers as
a function of radius has a minimum at intermediate radii in more
massive clusters like M3 \citep{fer93}, M55 (Zaggia, Piotto, \&
Capaccioli 1997), and 47 Tuc \citep{fer04}. The core population of
stragglers may be produced in collisions, while the envelope
population may be produced largely by primordial binary stars whose
orbits in the cluster have not changed significantly since the
cluster's formation \citep{map04}. Alternately, binaries that interact
within the core of the cluster can create blue stragglers with varying
amounts of recoil, so that the ones given little recoil quickly relax back to
the cluster core, while those given larger 
recoil end up spend mosting of their lives in the outskirts of
the cluster (Sigurdsson, Davies, \& Bolte 1994).

The continuing interest in blue stragglers is partly because they
may provide a means of gauging the recent dynamical history
of a cluster. Binary star systems are an important source of kinetic
energy that can be tapped by cluster stars during strong gravitational
interactions, which can in turn have a significant effect on the
evolution of the cluster's structure. The radial distributions of blue
stragglers may be revealing information about the timescale on which
dynamical friction modifies the stellar orbits \citep{map04}.  CMD
information may also lead to an understanding of blue straggler
formation \citep{sil99}, which may be connected to the dynamical
evolution of the cluster --- for example, 47 Tuc may have ceased
vigorous formation of blue stragglers several Gyr ago \citep{sil00}.

There are several reasons for focusing attention on low-density
clusters like Palomar 13. Observationally, it is a considerably easier
task to obtain a complete sample of good blue straggler candidates in
low-density clusters than in higher-density ones because of the greatly
reduced probability of image blending. Studies by \citet{fer95} and
\citet{pio04} have indicated that low-density globular clusters often
have some of the highest specific frequencies of blue stragglers. In
low density systems like Palomar 13, the expectation is that
primordial binary stars are most important in blue
straggler formation, although theoretical studies of open clusters
like M67 indicate that binary-binary interactions may still play an
important (or even dominant) role in producing stragglers
\citep{hur01}. There is good evidence that at least three blue
stragglers in M67 were formed via collisions involving more than two
stars: S977 \citep{leo96}, S 1082 \citep{vdb01,san03}, and S 1284
(based on the eccentricity of the short period orbit of the blue
straggler's companion; Milone \& Latham 1992). Other stragglers in
long-period eccentric binaries might also be the result of collisions
involving a binary system. For clusters that are on the verge of tidal
destruction, there may be a competition between decreasing stellar
density (due to the evaporation of low-mass stars) and
increasing binary fraction (resulting from mass segregation).

\subsection{Palomar 13 Blue Stragglers}

\begin{figure}
\plotone{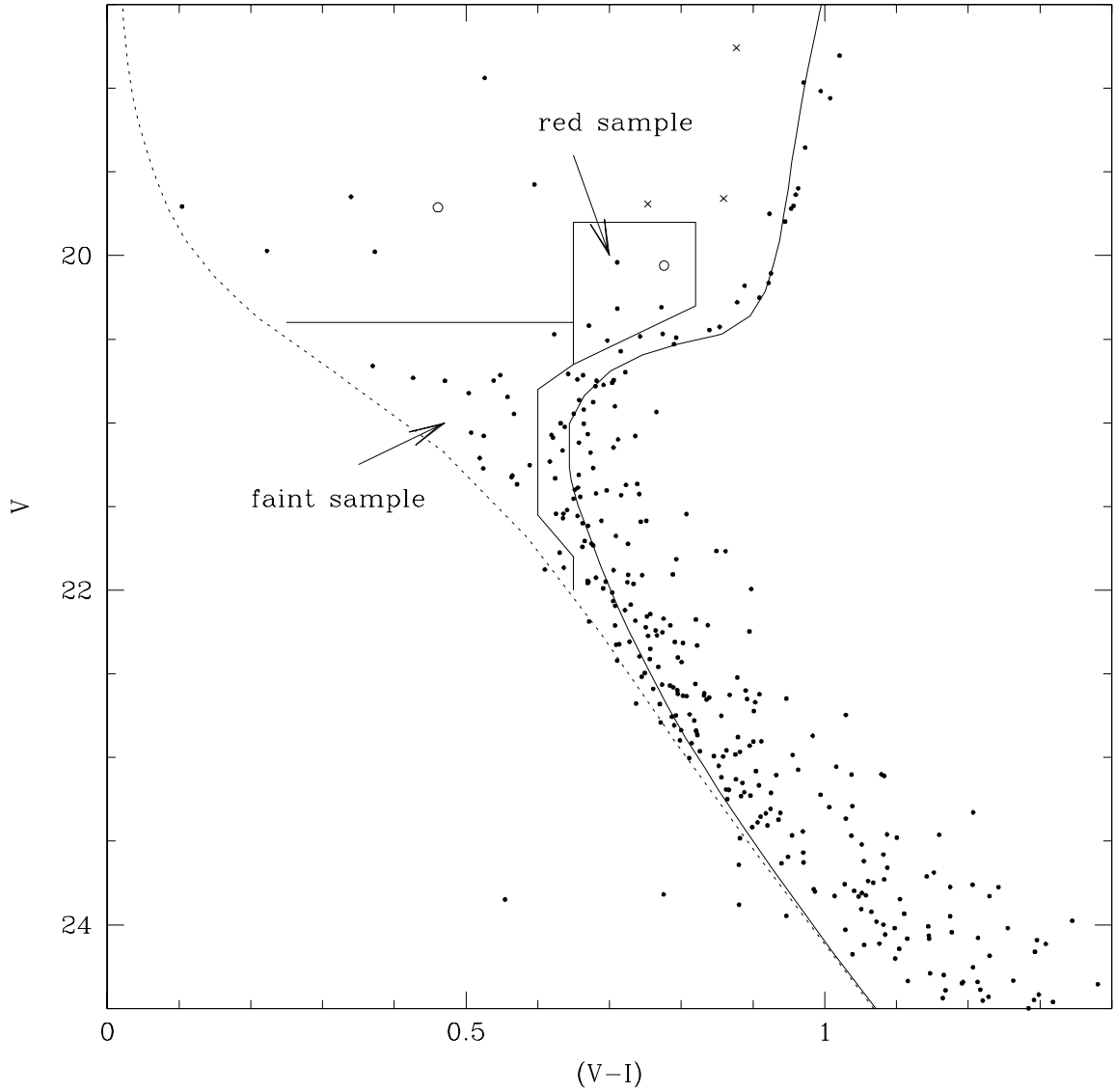}
\caption{Palomar 13 blue straggler star candidates: $\bigcirc$ show
candidates not found in our Keck field, and $\times$ show
high-probability field stars.  The {\it solid line} shows an 11.2 Gyr
isochrone and the {\it dashed line} shows 0.31 Gyr isochrone
\citep{gir02} for $Z = 0.0004$, E$(V-I) = 0.13$, and $(m - M)_V =
17.37$
\label{fig8}}
\end{figure}

As can be seen in the CMD (Fig. \ref{fig8}), Palomar 13 has a healthy
population of blue straggler stars, even outnumbering the horizontal
branch stars in the cluster. Blue stragglers are typically identified
as stars that are significantly brighter and bluer than than the
cluster's turnoff ($V < 21.2$). The region populated by stragglers is
roughly bounded on the blue side by a zero-age main sequence of the
appropriate composition. The reason for this is thought to be that
blue stragglers are rejuvenated stars with higher central hydrogen
content than single stars at the cluster turnoff. (This is independent
of the actual physical mechanism that causes this increase in central
hydrogen abundance.) In the open cluster M67 \citep{san03b}, there is
a large population of stars with colors between the fiducial line of
the cluster and the zero-age main sequence, but with brightnesses
placing them at or fainter than the cluster turnoff. These stars are
more difficult to identify in CMDs than canonical blue stragglers if
the photometry is poor, but there is every reason to identify them as
``rejuvenated'' blue stragglers. As a result, we have attempted to
identify {\it all} bright cluster members that have photometry placing
them significantly to the blue of the single-star sequence. This includes
three stars about 0.6 mag fainter than the turnoff that were identified 
by the significance of their deviation from the cluster fiducial line.

``Yellow'' or ``red stragglers'' are stars with colors between that of
the turnoff and the red giant branch, but brighter than the subgiant
branch. Such stars have been identified in open and globular star
clusters --- these stars may be former blue straggler stars that are
now evolving toward the giant branch, binary star systems or blends of
unassociated stars (most likely for stars within 0.75 mag of the
subgiant branch), or something more exotic. In a sparse globular
cluster like Pal 13, blends of unassociated stars are fairly
unlikely. For these reasons, we have also tried to identify stars that
could fall in this class.

Table \ref{tab5} lists all of the stars we identified (a total of 32)
as blue or red straggler star (BSS or RSS) candidates in Palomar 13 based
on position in the CMD and proper motions (Fig. \ref{fig8}). We also
cross-identified our list with the photometry of \citet{cot02} and
\citet{sie01}. Column 1 of the table gives the ID number from this
study, column 2 the ID number from \citet{sie01}, column 3 the ID
number from \citet{cot02}, columns 4 and 5 the position relative to
the cluster center in arcsec, columns 5-8 the $V$ magnitude and
$(V-I)$, $(B-V)$, and $(U-V)$ colors from the different datasets, and
column 9 the Siegel et al. membership probability. Star positions were
measured relative to the coordinates of the cluster center ($\alpha
= 23^{\mbox{h}}06^{\mbox{m}}44\fs8, \delta = 12\degr 46\arcmin 18\arcsec$;
epoch 2000.0) determined by \citet{sie01}. Proper motions from
\citet{sie01} were used to eliminate high-probability field stars when
possible. Two candidate BSS stars (IDs 67 and 117) are probable nonmembers ($0
< P_{\mu} < 0.4$), although one is at the faint end of the SMCT sample.
10 candidates have questionable membership ($0.4 <
P_{\mu} < 0.8$), and there is no proper motion information on another
five of our candidates. Since the field of \citet{cot02} extended
beyond our observed field, we used their photometry to identify 5
additional candidates (although three of 
these appear to be field stars).  Several of the stars in our list
fall close enough to the subgiant branch that there is a substantial
probability that they are detached binary systems --- these stars are
identified in the table, and were noted by SMCT as an apparent
sequence of stars parallel to the subgiant branch.

Borrisova et al. (1997) and \citet{sie01} each identified seven BSS
candidates. However, our smaller measurement errors for stars near the
cluster turnoff allow us to identify candidates that are closer to the
main-sequence turnoff in the CMD, and our better image resolution
allowed us to locate BSS candidates closer to the cluster core.  One
of the BSS candidates (BSS 4; SMCT ID 428) that was originally
identified by Borissova et al. (1997) is not included here because it
clearly falls within the cluster main-sequence turnoff band in our
photometry and has a low membership probability ($P = 0.23$). The
following subsections will provide analysis of our BSS population and
comparison of Palomar 13's population with other clusters.

The brightest straggler (ID 2) deserves some comment. From proper
motions, the star has a moderate probability of membership, but
\citet{blech} made two radial velocity measurements for the star
(25.85 and 25.67 km s$^{-1}$) placing it quite near the cluster mean
($24.1 \pm 0.5$ km s$^{-1}$). Given that it is the brightest blue
straggler in the cluster, its red color is surprising.  A
blend of a bluer blue straggler with a faint red giant can probably be
ruled out since the star's $U-V$ color is consistent with
colors using other filter combinations. Models of stellar collisions
(e.g. Sills \& Bailyn 1999) generally predict that the most massive
blue stragglers will tend to be among the bluest, primarily as a
result of the relationship between effective temperature and mass for
main sequence stars. Spectroscopic measurement of the star's surface
gravity might help verify whether it is an evolved blue straggler or
not.

At the risk of overanalyzing this relatively small population of
stragglers, we note a couple of features of the CMD distribution.  The
color distribution for the seven brightest BSS shows no
tendency for the stragglers to cluster toward the zero-age main
sequence. In fact, these stragglers are spread almost evenly from near
the zero-age main sequence to the turnoff color. This type of color
distribution is seen in many other clusters, and is not predicted by
models of stellar collisions (e.g., Sills \& Bailyn 1999). In
addition, there is a gap of over 0.5 mag between the bright blue
stragglers and the 17 straggler candidates fainter than the
subgiant branch.

\subsection{Radial Distribution} \label{4:two}

\begin{figure}
\plotone{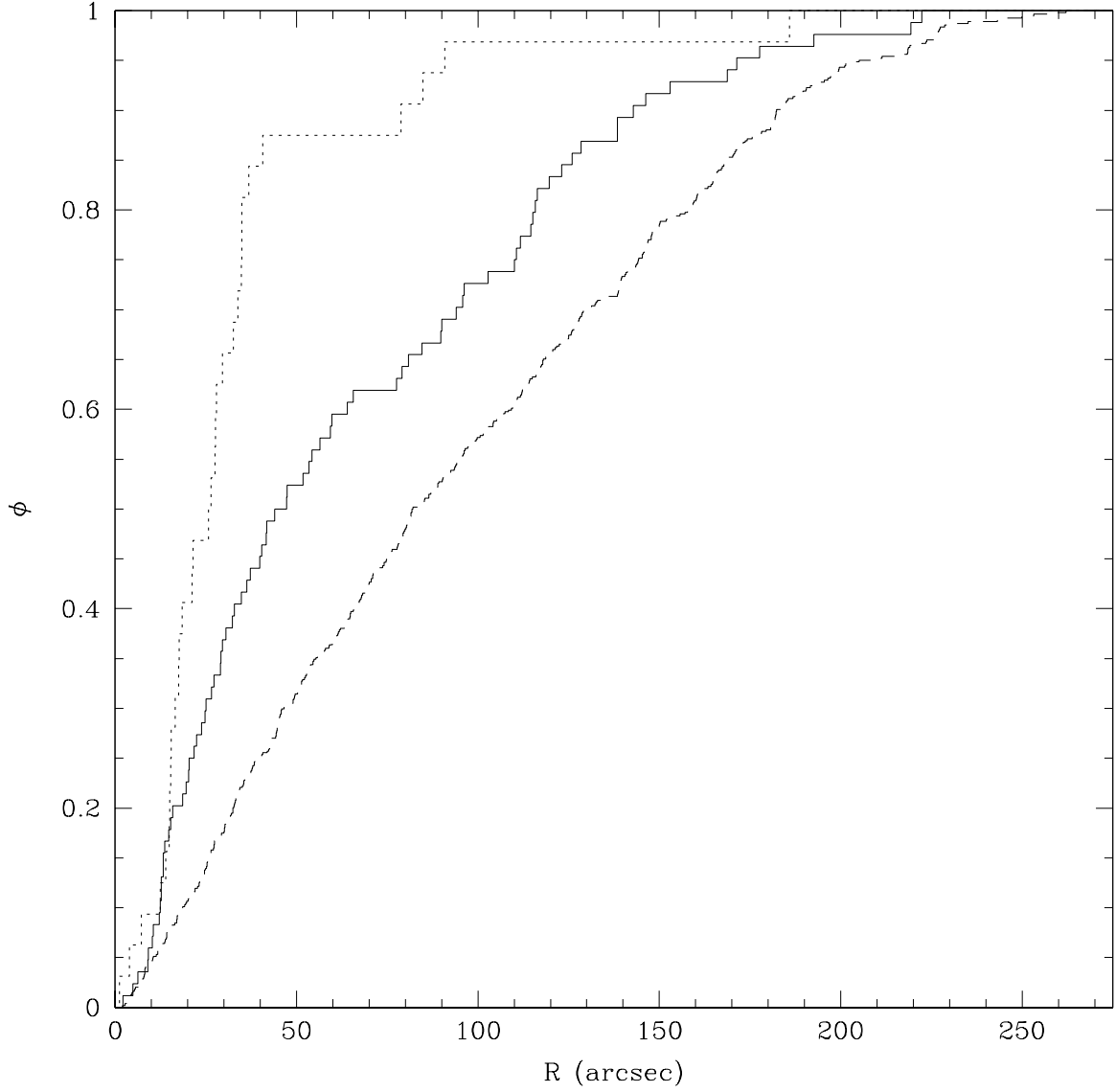}
\caption{Normalized cumulative radial distributions for giant and subgiant
stars ({\it solid line}), main sequence stars ({\it dashed line}), and blue
stragglers ({\it dotted line}).\label{fig9}}
\end{figure}

In an effort to learn more about the effects of cluster dynamics on
the BSS population, we compared the cumulative radial distribution of
the BSS to the populations of giant branch ($V < V_{TO}$) and main
sequence stars. In order to cover the widest range of radii possible,
we combined our data with those of \citet{cot02}. Field stars were
eliminated using proper motion \citep{sie01} and radial velocity
\citep{cot02} information for the giant stars, and CMD location for
the main-sequence stars. After these cuts, there were 83 stars in the
giant sample and 543 stars in the main sequence sample. Typically the
straggler population is compared to brighter populations (for example,
horizontal branch stars), but because these stars are so rare in
Palomar 13, we were forced to resort to fainter
populations. Kolmogorov-Smirnov (K-S) probability tests were used to
test the hypothesis that the populations were drawn from the same
parent population. The cumulative radial distributions are shown in
Fig. \ref{fig9}.  The results of the K-S tests are given in Table
\ref{tabks}, including the absolute deviation $D$ and the probability
$P$ that the two samples were drawn from the same distribution.

The central concentration of the blue stragglers relative to the
cluster light has been noted before by \citet{sie01}. However, our
results show that there are {\it significant} differences between all
three samples, with the BSS distribution being the most centrally
concentrated, and the MS sample being the least concentrated. The
probability that any one of the populations was drawn from the same
distribution as another is less than 0.06\%.  Just four of the
selected BSS are found outside a radius of $42\arcsec$ ($\sim 3 r_c$
according to \citet{cot02}).

Because the number of blue stragglers is relatively small, we now
examine what effect the exclusion of subsets of stragglers has on
the results of the K-S tests. The elimination of the six BSS candidates
redder than the turnoff color (stars which have a considerably higher
probability of being detached binaries) has a minor effect on
our conclusions: there is still a probability of less than 0.1\% that the
BSS and giant samples come from the same distribution. We have also
looked at the distributions of the brightest 7 and faintest 19 of the
stragglers and find that their radial distributions are
indistinguishable.  There is a 12\% chance that the bright BSS sample
was drawn from the same distribution as the giants, although this is
primarily due to the small size of the sample and the single bright BSS
(BSS 7 from SMCT) at $185\arcsec$ from the cluster
center. The faint sample has just a 0.4\% chance of being drawn from
the same distribution as the giants.

\begin{figure}
\plotone{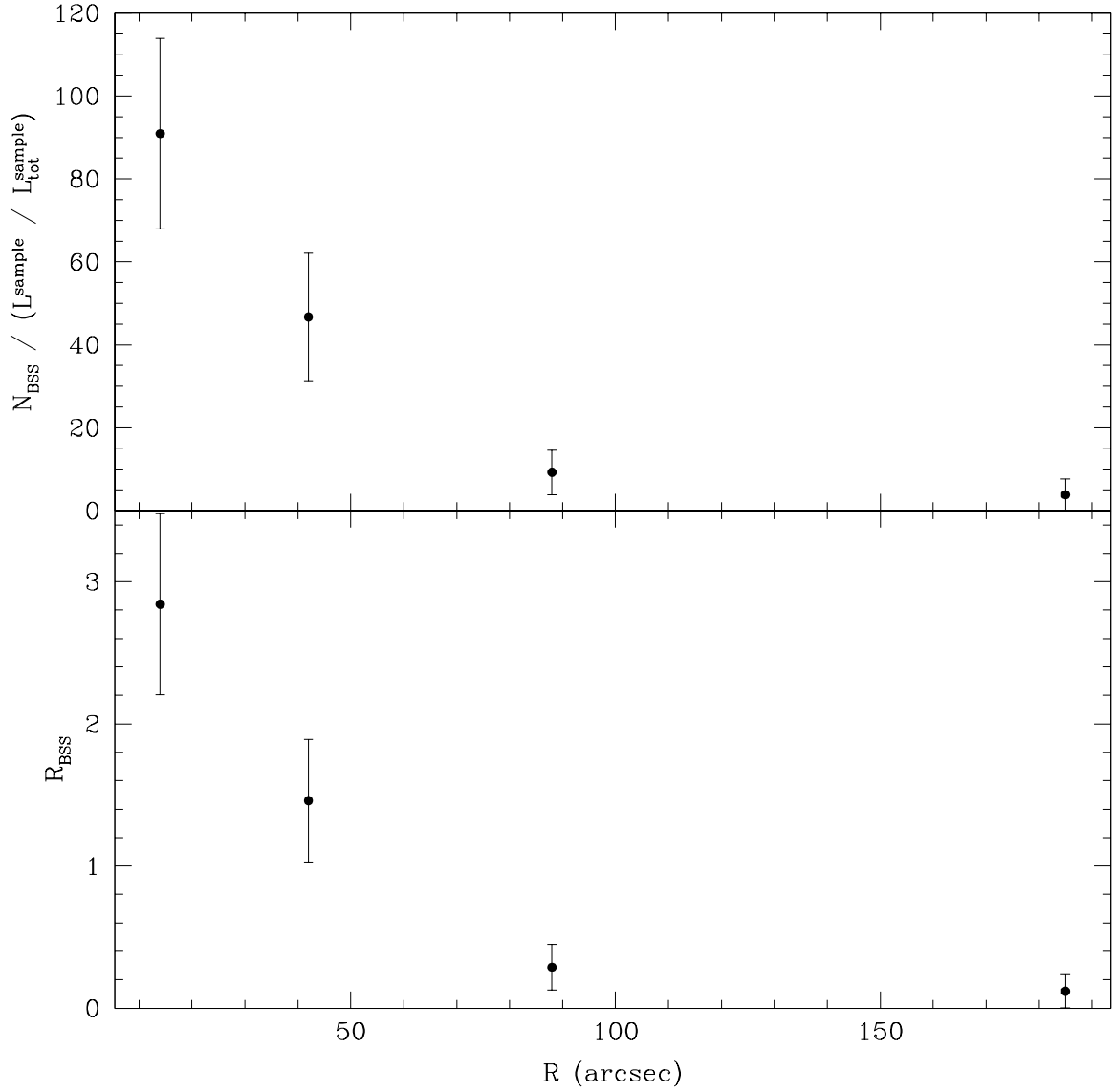}
\caption{{\it Top panel:} Frequency of blue stragglers relative to the
integrated $V$-band flux of detected cluster stars.  {\it Bottom
panel:} specific frequency of blue stragglers relative to the
integrated flux.\label{figr}}
\end{figure}

The cumulative radial distributions should make it clear that there is
a large increase in the relative frequency of blue stragglers going toward the core
of the cluster. Recent studies of much more massive globular clusters
\citep{fer93,zag97,fer04} have indicated that the blue straggler
frequency decreases at intermediate radii before rising again at large
distances from the center. Because the typical normalizing populations
(HB or bright RGB stars) are small in number in Palomar 13, we have chosen
to normalize the blue straggler frequency to the fraction of the
integrated flux from all other detected cluster stars (see
Fig. \ref{figr}).

We also plot the specific frequency defined by \citet{fer93}. We find
that the specific frequency of blue stragglers in the core of Palomar
13 is comparable to that of much denser clusters like M3 \citep{fer93}
and 47 Tuc \citep{fer04}, but we see no signs of an upturn in the
specific frequency of stragglers at large radii seen in these denser
clusters. In M3 and 47 Tuc, this upturn occurs at $\sim 10 r_c$.  In
the low-concentration ($c = 0.8$) cluster M55, the upturn occurs at $r
\simeq 2 r_c$ \citep{zag97}.  Given $r_c = 14\arcsec$ \citep{cot02},
Palomar 13 has been surveyed out to nearly $18 r_c$. The lack of an
upturn probably cannot be ascribed to small number statistics: the
specific frequency would need to be almost 10 times higher than
measured to be consistent with the denser clusters. However, there is
a possibility that the relative frequency of stragglers could increase
at much larger radii than surveyed here given that the best fitting
King-Michie model has a tidal radius $r_t = 26\arcmin \pm 4\arcmin$
\citep{cot02}. From the wider-field ($7\arcmin \times 14\arcmin$) CFHT
photometry of \citet{cot02} (their Fig. 6), there is only one blue
straggler candidate outside the field discussed in this paper, and it
is at more than twice the distance from cluster center as any other
straggler and has no membership information at present. While this
result should be tested in surveys of other low-mass globular
clusters, the blue straggler distribution may be fundamentally
different than those of more massive clusters.


\subsection{Blue Straggler Population Comparison} \label{4:three} 

To put the Palomar 13 blue straggler population in perspective, we would
like to compare the global BSS population to that of other globular
clusters. Recently Piotto et al. (2004) published results for
frequencies of blue stragglers relative to HB and RGB stars for 56
globular clusters.  Palomar 13 has much lower values for integrated luminosity and
central luminosity density than any of the clusters in the
Piotto et al. sample.

Piotto et al. found a significant anticorrelation between the specific
frequency $F_{BSS}^{HB} = N_{BSS} / N_{HB}$ and integrated luminosity
$M_V$, a weaker anticorrelation with central density $\log \rho_0$,
and no significant trend with expected collision rate $\log
\Gamma_{\star}$. From the globular cluster catalog of Harris (1996),
Palomar 13 has $M_V = -3.74$, $\log \rho_0 = 0.40$ ($\rho_0$ in $L_{\odot}
\: \mbox{pc}^{-3}$), and $\log \Gamma_{\star} = -16.25$ (calculated
according to the procedure given in Piotto et al.). There are five HB
stars (four RR Lyrae stars, and one nonvariable) known in the cluster
within the area covered by our photometry and that of
\citet{cot02}. The blue straggler samples in \citet{pio04} were not
selected using uniform criteria for all clusters (F. De Angeli,
priv. comm.), so we discuss two selections. Restricting our blue
straggler sample to those having $V < 20.4$ (the level of the base of
the giant branch) and $(V-I) < 0.65$, we are left with 7 stars, giving
$F_{BSS}^{HB} = 1.4 \pm 0.8$ (using Poisson error estimates). Only two
clusters (NGC 6717 and NGC 6838) in the Piotto et al.  study have
higher values. These two clusters were the faintest ones included in
the Piotto et al. study, though they have central densities that are
more than two orders of magnitude higher.  If we accept all of the
blue straggler candidates that are bluer than the turnoff, we have
$N_{BSS} = 26$ and $F_{BSS}^{HB} = 5.2 \pm 2.5$, which is higher than
any cluster value in the Piotto et al. sample.

Piotto et al. also examined the blue straggler population relative to
the amount of flux sampled, and once again they found a significant
anticorrelation with the absolute magnitude of the cluster. When the
\citet{cot02} dataset is combined with ours, nearly the entire flux from the
cluster is sampled, which gives us $\log N_S(BSS) = -0.65$ or $-0.08$
depending on whether the more restrictive sample of blue stragglers is
used. (For comparison, $\log N_S(HB) = -0.80$.) For the
restricted sample, the $\log N_S(BSS)$ value places Palomar 13 among the
clusters with the highest values, but for the total sample, Palomar 13
has a value second only to NGC 6838, a cluster that is more than 5
times brighter ($M_V = -5.60$; Harris 1996).

This study makes Palomar 13 the faintest globular cluster with a
thoroughly studied BSS population, covering from the cluster center
out to nearly 18 core radii. In combination with the results from
\citet{pio04}, Palomar 13 provides additional evidence that the
production of blue stragglers relative to the cluster HB stars or the
cluster light may plateau for clusters with $M_V \gtrsim -6$ (see Fig. 1
of \citet{pio04}).  On the other hand, because Palomar 13 appears to
be a cluster in the throes of tidal disruption by the Galaxy, it is
possible that the binary fraction in the cluster may have been
artificially elevated during mass segregation, affecting the size of
the straggler population.

\section{Binary Fraction} \label{binary}

\begin{figure}
\plotone{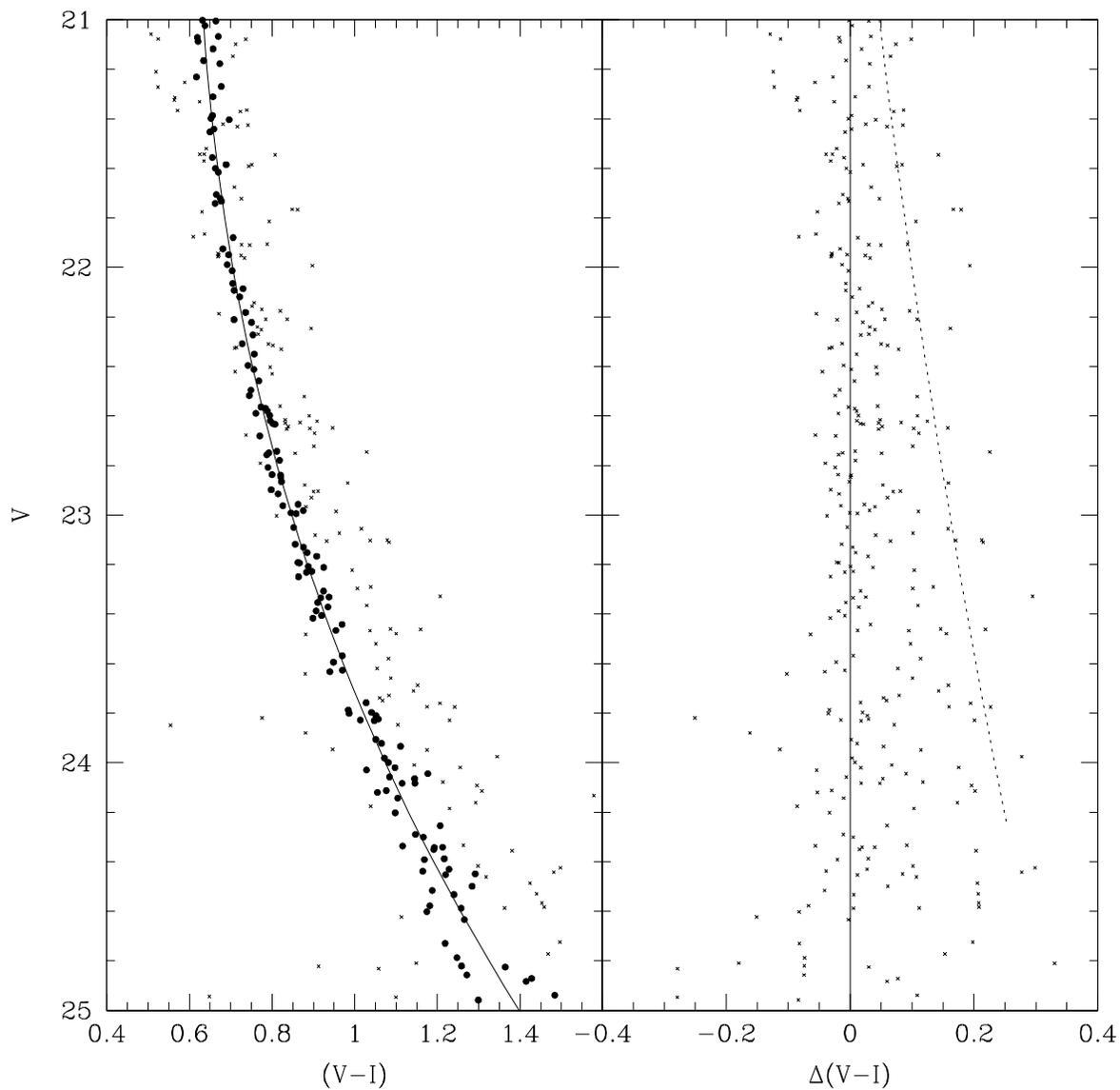}
\caption{{\it Left panel:} The Palomar 13 main sequence CMD. The {\it
solid line} is the polynomial fit to the main sequence stars shown
with $\bullet$.  These stars have photometric errors placing them less
than $2.5 \sigma$ from the polynomial fit. {\it Right panel:} The
rectified CMD (with the color of the polynomial fit subtracted from
the color of each star). The {\it dotted line} shows the expected positions
for equal-mass binaries. \label{figrect}}
\end{figure}

A characterization of the binary population in Palomar 13 can provide
important information about the dynamical environment and history of
the cluster. As a cluster undergoes mass segregation as a result of
internal dynamics, and loses stars due to its tidal interaction with
the Galaxy, the fraction of cluster stars in multiple star systems
will increase. Siegel et al. (2001) noticed an apparent double
subgiant branch that could be due to the presence of near equal-mass
binary stars. Their photometry only reached the main sequence turnoff,
however. \citet{cot02} noticed evidence for a ``second sequence''
running parallel to the cluster's main sequence in its CMD, but did
not characterize it. As can be seen in Fig. \ref{figrect}, the
sequence is better defined in our dataset.

Probably the most direct means of estimating the global binary
fraction is to characterize the population of cluster stars falling to
the red of the single-star main sequence \citep{rub97, bol92, rom91},
where binary stars and optical blends of cluster stars fall in the
CMD. We have chosen to use a method similar to the one outlined by
Bolte (1992) that provides an estimate of the fraction of stars in a
cluster that are in binary systems, independent of assumptions about
binary star properties like the mass ratio distribution.

Our analysis only makes use of photometry from this study since the
sample of \citet{cot02} had larger amounts of photometric scatter and
field star contamination.  The main sequence stars in the CMD were
first used to create a fiducial line for the cluster
(Fig. \ref{figrect}). This was accomplished by iteratively fitting the
main sequence sample with a polynomial and rejecting stars with
photometry that placed them more than a threshold number of $\sigma$
(from the photometric errors) away from the fit. The threshold value
was initially chosen to be large ($10 \sigma$) and was then
reduced. The final fit used a fifth-order polynomial and a $2.5 \sigma$
rejection cut. Because the stars toward the blue side of the main
sequence are most common, the fit naturally migrates toward that edge
(where the single-star main sequence is expected to be).  This method
also has the advantage of requiring minimal human intervention and no
judgement on which stars to select for the fit.

\begin{figure}
\plotone{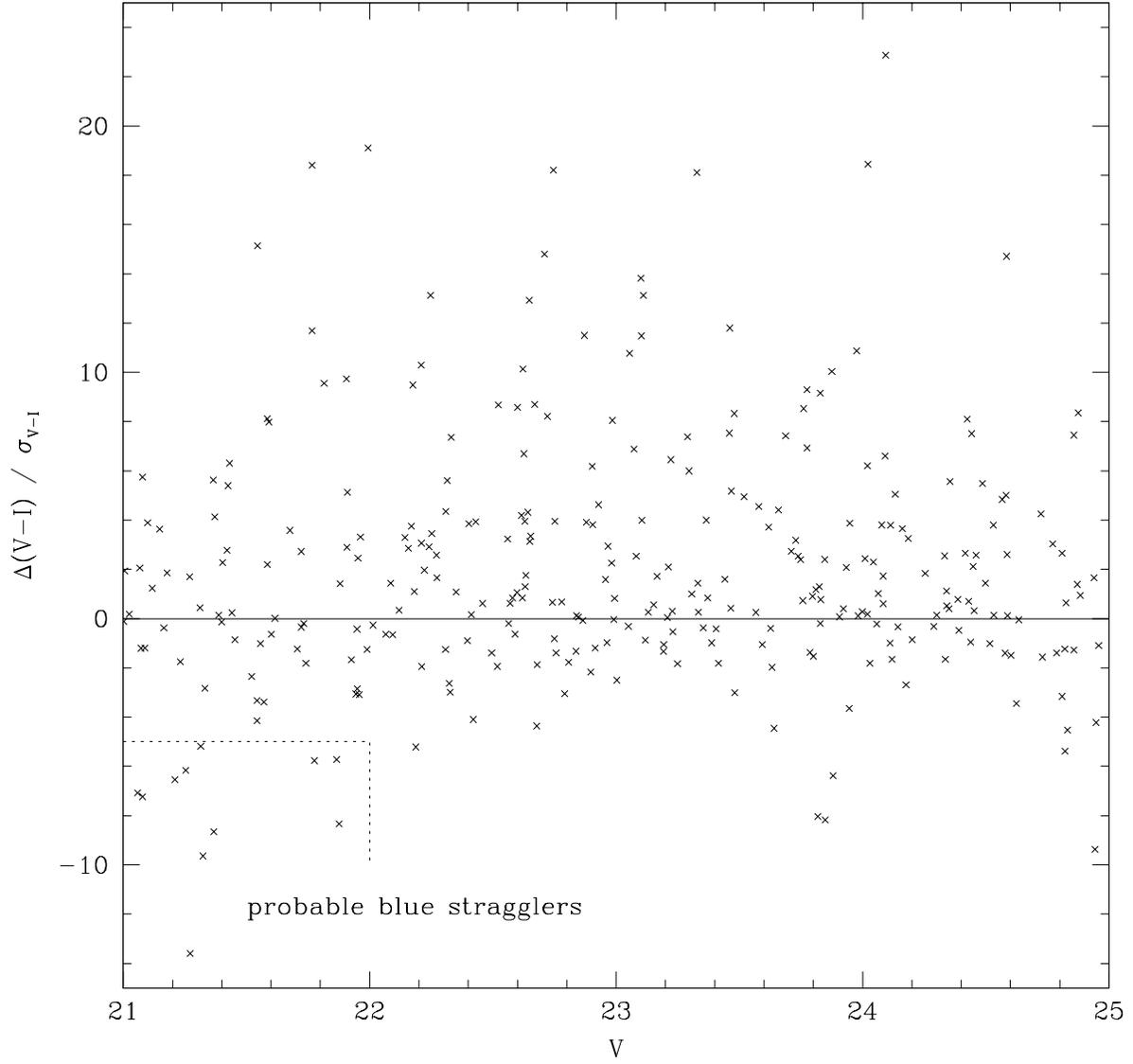}
\caption{The significance of color deviations $\Delta (V-I) / \sigma_{V-I}$
versus magnitude for Palomar 13 stars. Positive $\Delta (V-I)$ indicates
that the star is to the red of the main sequence fit.\label{signif}}
\end{figure}

We then determined the color deviation $\Delta (V-I)$ between the
color of each star and the color of the main sequence fit. Thus,
single stars will be distributed about $\Delta (V-I) = 0$, while
main sequence binary stars will be distributed to the red ($\Delta
(V-I) > 0$).  The middle panel of Fig. \ref{figrect} shows a line
corresponding to the expected positions of equal-mass binary systems
--- there is a clear excess of stars found to the red of the main
sequence as compared to the blue side. Fig. \ref{signif} plots the
number of $\sigma_{V-I}$ of deviation from the main sequence fit
versus magnitude, showing that the stars to the red have significant deviations.

\begin{figure}
\plotone{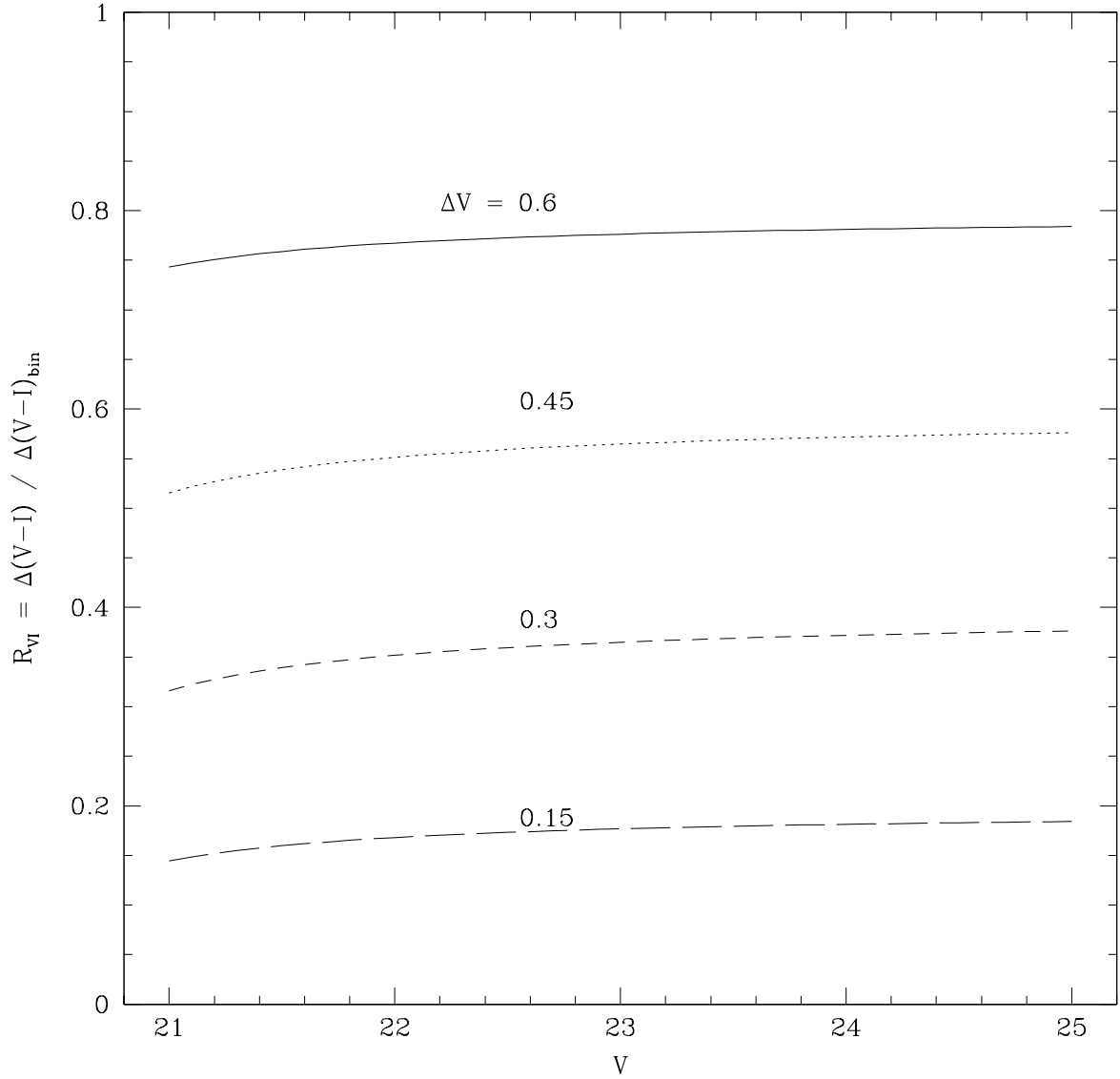}
\caption{The relationship between $R_{VI} = \Delta (V-I) / \Delta
(V-I)_{bin}$ and the $\Delta V$ (the magnitude change to a star's
measured magnitude due to the addition of the light of an unresolved
star). \label{figrvi}}
\end{figure}

The derivation of a binary
fraction from this distribution is complicated by the small number of
stars in the cluster and by changes in the slope of the main sequence
as a function of magnitude. Therefore, we opted to divide the color
deviations for each star by the {\it expected} color deviation for an
equal-mass binary system $\Delta (V-I)_{bin}$ to get the quantity
$R_{VI} = \Delta (V-I) / \Delta (V-I)_{bin}$. This has the benefits
that there is a monotonic relationship between $R_{VI}$ and $\Delta V$
(the difference between the magnitude of the primary and the magnitude
of the binary system), and $R_{VI}$ is very nearly constant for a
binary system of given $\Delta V$ along the entire main sequence (see
Fig. \ref{figrvi}). The result of this procedure is shown in the right
panel of Fig. \ref{figrect}. In this way, stars measured over most of
the magnitude range of the observed main sequence can be
compared. This procedure amplifies measurement errors on the upper
main sequence (because the main sequence is steeper, and $\Delta
(V-I)_{bin}$ is smaller). However, the photometric errors for these
stars are usually much smaller than for stars on the lower main
sequence, so that this amplification is a minor effect.

The distribution of stars in $R_{VI} = \Delta (V-I) / \Delta
(V-I)_{bin}$ was then histogrammed for $21.2 < V
< 24.2$ (eliminating faint stars with the largest photometric
uncertainties and stars above the level of the turnoff). The
estimation of the binary fraction is dependent on the assumption that
the single star population of the cluster falls in a Gaussian
distribution centered around $R_{VI} = 0$. We estimated the size of the
dominant population of single stars and binary stars with extreme
luminosity ratios ($R_{VI} \approx 0$) with two different
procedures.  

\begin{figure}
\plotone{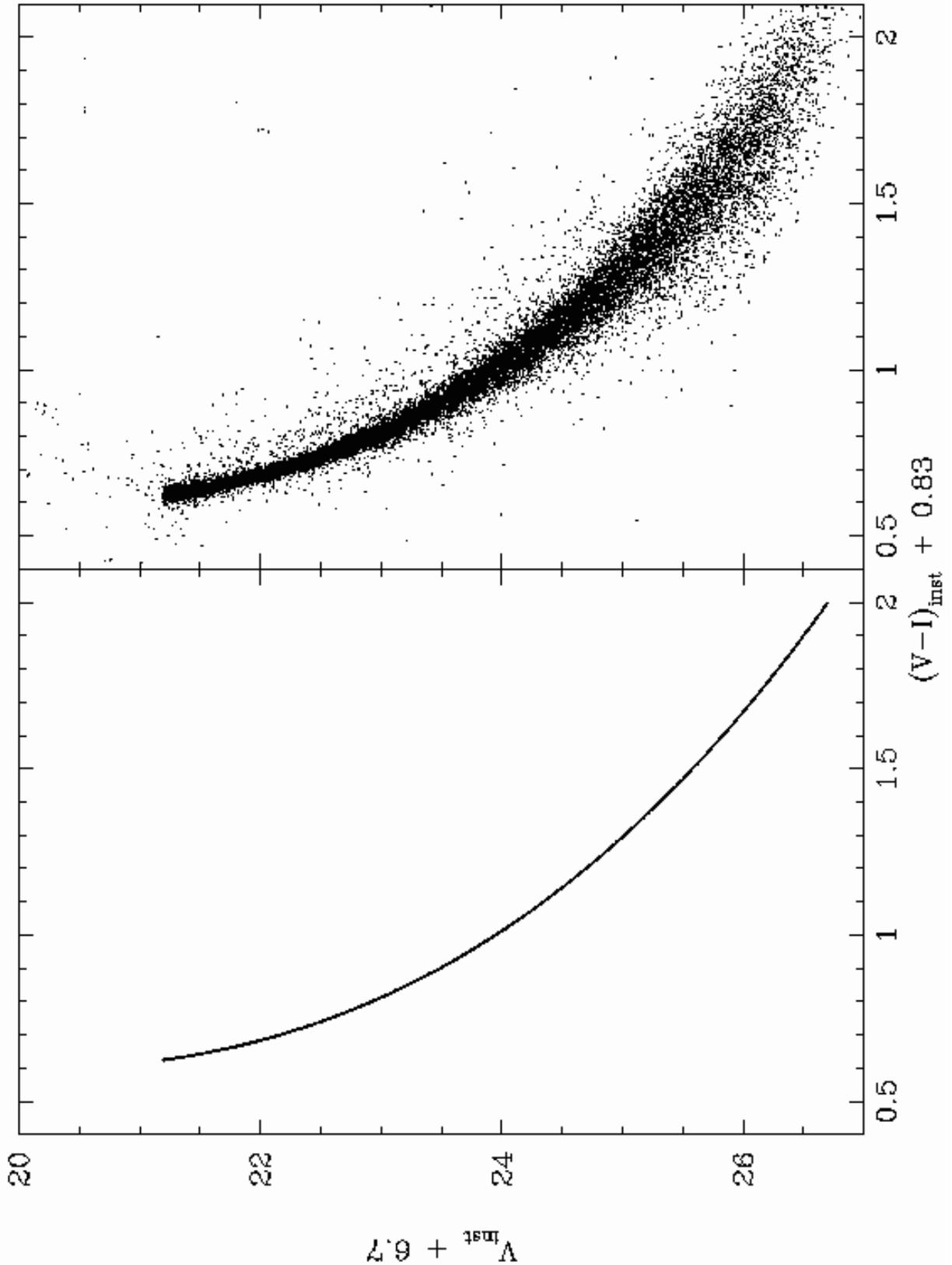}
\caption{Color-magnitude diagrams of artificial star experiments on
the Keck ESI data. The instrumental magnitude and color have been
adjusted to roughly overlie the calibrated data. {\it Left panel:}
Input artificial stars. {\it Right panel:} Photometry of objects
detected at the positions of the input artificial stars.\label{artcmd}}
\end{figure}

Our first method employed artificial star tests to predict the
$R_{VI}$ distribution for a set of single stars covering a range of
magnitudes. This distribution will differ from Gaussian because
fainter stars have larger measurement errors and will therefore cover
a larger range in $R_{VI}$. We conducted 30 artificial star runs on
our images, placing approximately 700 artificial main sequence stars
in the field for each run. The CMDs of the input and recovered
artificial stars are shown in Fig. \ref{artcmd}. We then determined
the width of the distribution in $R_{VI}$ for all stars in 0.5 mag
bins by fitting a Gaussian to stars within $1.5 \sigma$ of $R_{VI} = 0$.
The predicted single-star distribution for real stars was then
computed by summing Gaussians of unit area (one Gaussian for each star
within $2 \sigma$ of $R_{VI} = 0$) having FWHMs corresponding to that
of the artificial star distribution for the same magnitude bin.  After
the predicted distribution was subtracted from the distribution of
real stars, the red ($R_{VI} > 0$) and blue sides of the residual
distribution were integrated. If the residuals on the blue side are
part of a symmetric distribution of stars with large photometric
errors, then an equal contribution must be subtracted from the red
side in order to compute the binary fraction.

Using this method on the distribution of real stars, we found 121.7
stars in the central peak, and once this contribution was subtracted,
there were 83.4 stars left on the red side of the peak, and 12.0 stars
on the blue side. (9 stars on the blue side were blue stragglers, and
were not considered.) This results in a net contribution of 71.4 stars
from the red side, which produces a binary fraction of $0.31 \pm 0.04$
(with the uncertainty computed using Poisson statistics).

\begin{figure}
\plotone{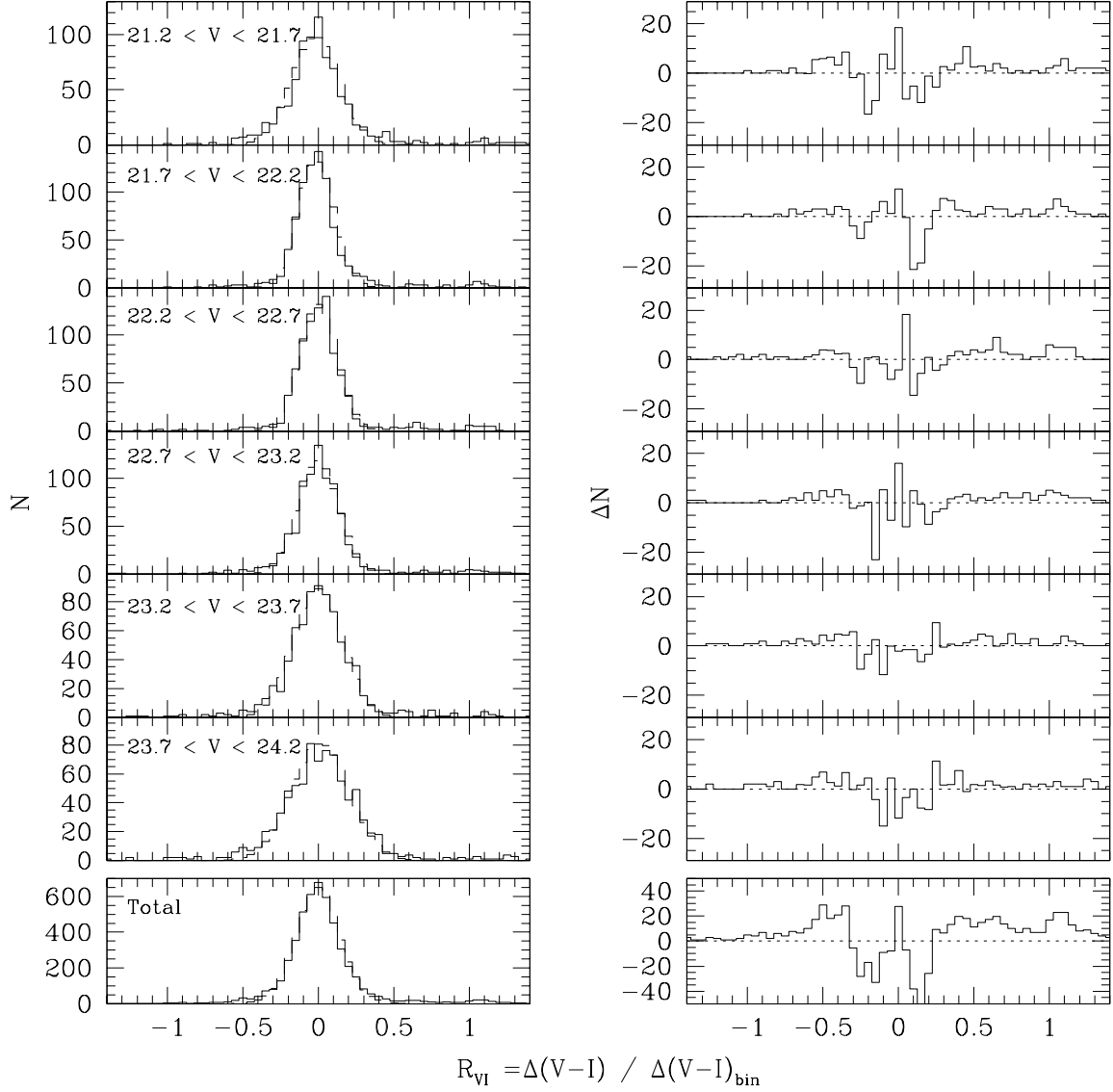}
\caption{{\it Left panels:} Normalized color deviations $R_{VI}$ for
artificial stars within $1\arcmin$ of the center of the cluster,
separated by magnitude. A Gaussian fit ({\it dashed line}) to the
distribution is shown in the top 6 panels, while the sum of these fits is shown in the ``Total'' panel. {\it Right panels:} The residuals after the
fits are subtracted from the distributions.\label{artfig}}
\end{figure}

As a check, we investigated whether unresolved blends of unassociated
single stars could produce the binary fraction measured above. We
determined the binary fraction that would be determined from all
artificial stars placed within $1\arcmin$ of the cluster center. The
result of these artificial star tests is shown in \ref{artfig}. For
the 4738 artificial stars falling in the magnitude range analyzed for
the binary fraction, we found 107 net stars on the blue side and 169 net
stars on the red side of the predicted single-star distribution,
giving a binary fraction $0.013 \pm 0.002$. Although the
artificial star experiments demonstrate that blends contribute to the
binary fraction measurement even in a cluster this sparse, blends are
far too unlikely to explain a binary fraction as high as 30\%.

\begin{figure}
\plotone{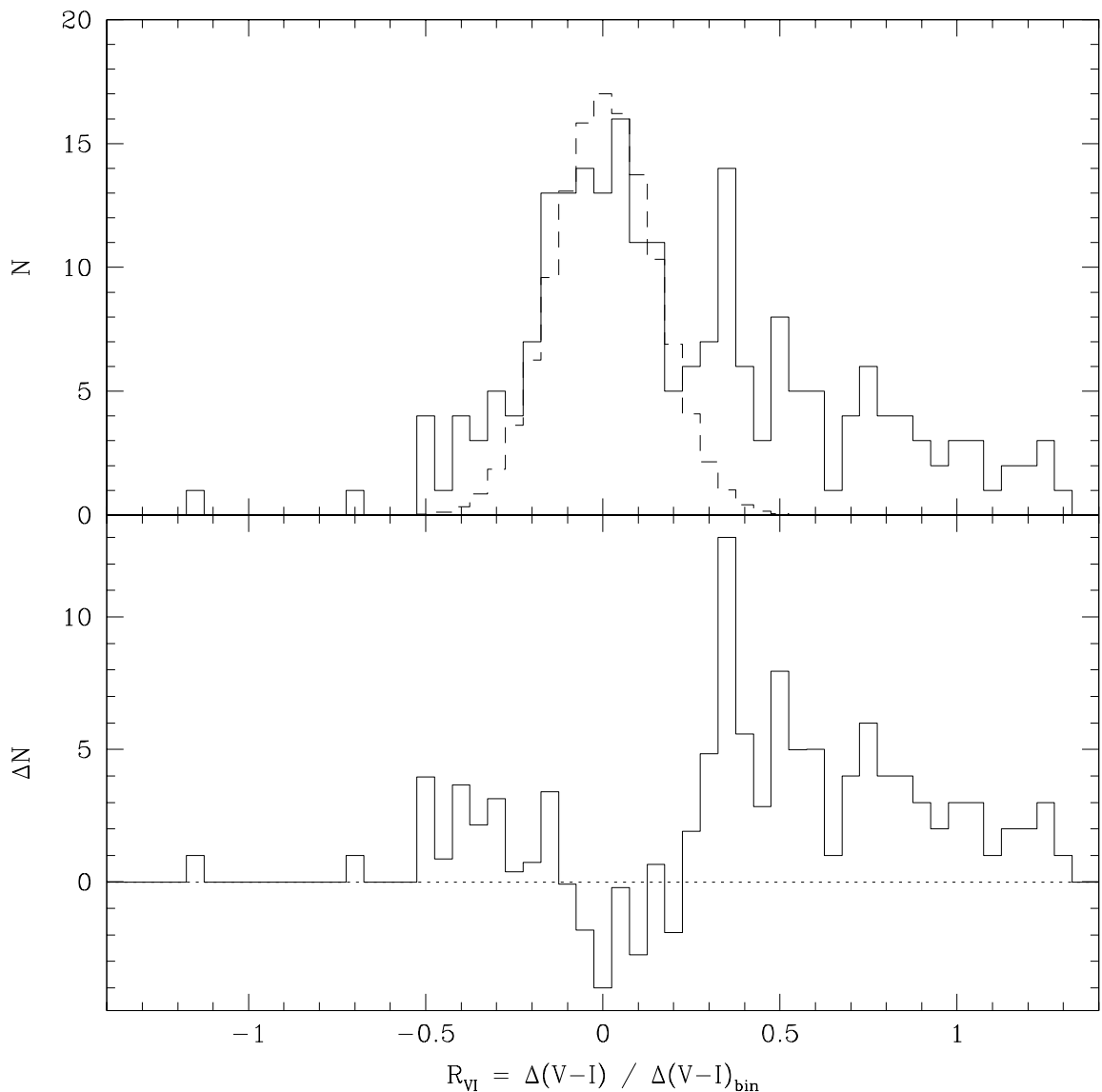}
\caption{{\it Top panel:} The distribution of color deviations (star
color minus the color of single star sequence at the same $V$
mangitude) relative to the color difference between the single star
sequence and equal-mass binary sequence for Palomar 13 main sequence
stars. The {\it dashed line} shows the Gaussian fit to stars with $\vert \Delta
(V-I) / \Delta (V-I)_{bin} \vert < 0.3$. {\it Bottom panel:} The subtraction of
the Gaussian fit from the star distribution.\label{fighist}}
\end{figure}

For the second method of determining the binary fraction, we simply
fit a Gaussian distribution to the central peak of the real star
distribution ($\vert R_{VI} \vert < 0.3$), returning a mean of
$0.0052$ and $\sigma = 0.144$ (see Fig. \ref{fighist}). After the Gaussian was
subtracted, the red and blue sides were integrated out to $R_{VI} =
1.4$.
We integrated beyond $R_{VI} = 1$ on the red side because errors in
the photometry of nearly equal-mass binary systems could cause them to
have $R_{VI} > 1$. The integrations of the blue and red sides returned
12.4 and 80.9 stars respectively. (Once again, blue
stragglers were not considered.) If we again assume that the
distribution on the blue side is part of a symmetric distribution of
stars with large photometric errors, we are left with a net
contribution of 68.5 stars from the red side.  We measured 123.7 stars
in the central peak. Adding in the contribution from the stars on the
blue side and an equal contribution from stars on the red side yields
a total single-star content of 148.5 stars, which results in a binary
fraction of $0.32 \pm 0.04$.  The results from the two methods
described above matched well in part because we restricted ourselves
to stars significantly brighter than the faint limit of the survey, so
that photometric errors did not end up broadening the total star
distribution excessively.

In the deep exposures we discussed here, we also need to consider
contaminating objects like background galaxies and field stars.  The
primary source of contamination for the upper main sequence seems to
be galaxies. Our use of a cut on image SHARP-ness eliminated a large
fraction of the objects red of the expected position of the equal-mass
binary sequence and blue of the single-star main sequence. Based on
this, we do not regard galaxies as a likely source of contamination
for the upper main sequence used in the binary fraction
analysis. Field star contamination could artificially boost the
measured binary fraction if a significant number of stars fall between
the single-star main sequence and the equal-mass binary sequence.  The
blue edge of the field star population falls near the main sequence of
Pal 13 (for example, compare the $BV$ data of the north Galactic pole
in Fig. 2a of \citet{reid} with Fig. 4 of \citet{cot02}). However, the
field star distribution extends far to the red. As a result, the very
small number of stars redward of the equal-mass binary sequence can be
taken as evidence that the field star contribution is minimal.

The general technique was first applied by \citet{bol92} to NGC 288,
giving $f_b \sim 0.15$. Comparing Bolte's Fig. 11 with our
Fig. \ref{fighist}, the binary fraction in Palomar 13 is noticeably
larger than in NGC 288. Veronesi et al. (1996) examined the loose
globular cluster E3 using a method roughly equivalent to ours and
found $f_{b} = 0.29 \pm 0.09$, which is quite similar to the value we
derive for Palomar 13. On the other hand, \citet{koch} found $f_b =
0.09 \pm 0.01$ for the core ($r < r_c$) of the cluster Pal 5, with
comparable values out to 3 core radii. Of these clusters, E3 and Pal
13 have similar integrated magnitudes ($M_{V_t} = -2.77$ and $-3.74$,
respectively; Harris 1996), while Pal 5 ($M_{V_t} = -4.77$;
\citet{oden}) and NGC 288 ($M_{V_t} = -6.74$; Harris 1996) are
considerably brighter. The idea that fainter clusters have higher binary 
fractions should tested with additional studies
of clusters in this range.

Because there is not a noticeable residual of stars with $0 < R_{VI}
< 0.3$, we can consider these measurements to be the binary fraction
for systems with $R_{VI} > 0.3$ ($R_{VI} = 0.3$ roughly corresponds to
$\Delta V = 0.25$, which means the primary contributes about 80\% of
the system light.) The numbers we quote above are rough {\it lower
limits} to the {\it total} percentage of unresolved binary systems
since systems with small mass ratios would tend to be counted as
single stars. An estimation of the total binary fraction requires
assumptions about the distribution of mass ratios in binary systems
$F(q)$ and extensive artificial star tests. Such studies have been
carried out for the globular clusters NGC 288 ($0.08 \la f_{b} \la
0.38$ within 1.6 core radii, and $f_b < 0.1$ outside that; Bellazzini
et al. 2002) and NGC 6752 ($0.15 \la f_{b} \la 0.38$ for the inner
core radius, $f_{b} \leq 0.16$ for the outer core radius; Rubenstein
\& Bailyn 1997). These two clusters have central densities that differ
by a factor of $\sim 1300$ (Harris 1996), and probably also have very
different dynamical histories. \citet{fer03a} find evidence that the
surface brightness profile of NGC 6752 is best fitted by a {\it
double} King-model profile, concluding that it is in a
post-core-collapse phase. NGC 288 has probably had a history similar
to that of Palomar 13, since there are indications of an extratidal
tail implying significant mass loss from the cluster
\citep{for01}. However, in their detailed analysis of NGC 288,
Bellazzini et al. found that although they achieved compatibility
between their numerical simulations and observations for $0.05 \leq
f_b \leq 0.30$, the highest degree of compatibility was for $0.10 \leq
f_b \leq 0.20$ (in agreement with Bolte's earlier analysis). We
conclude that it is very likely that Palomar 13 has a higher binary
fraction than NGC 288, and is almost certainly in a more advanced
stage in its tidal destruction.

\begin{figure}
\plotone{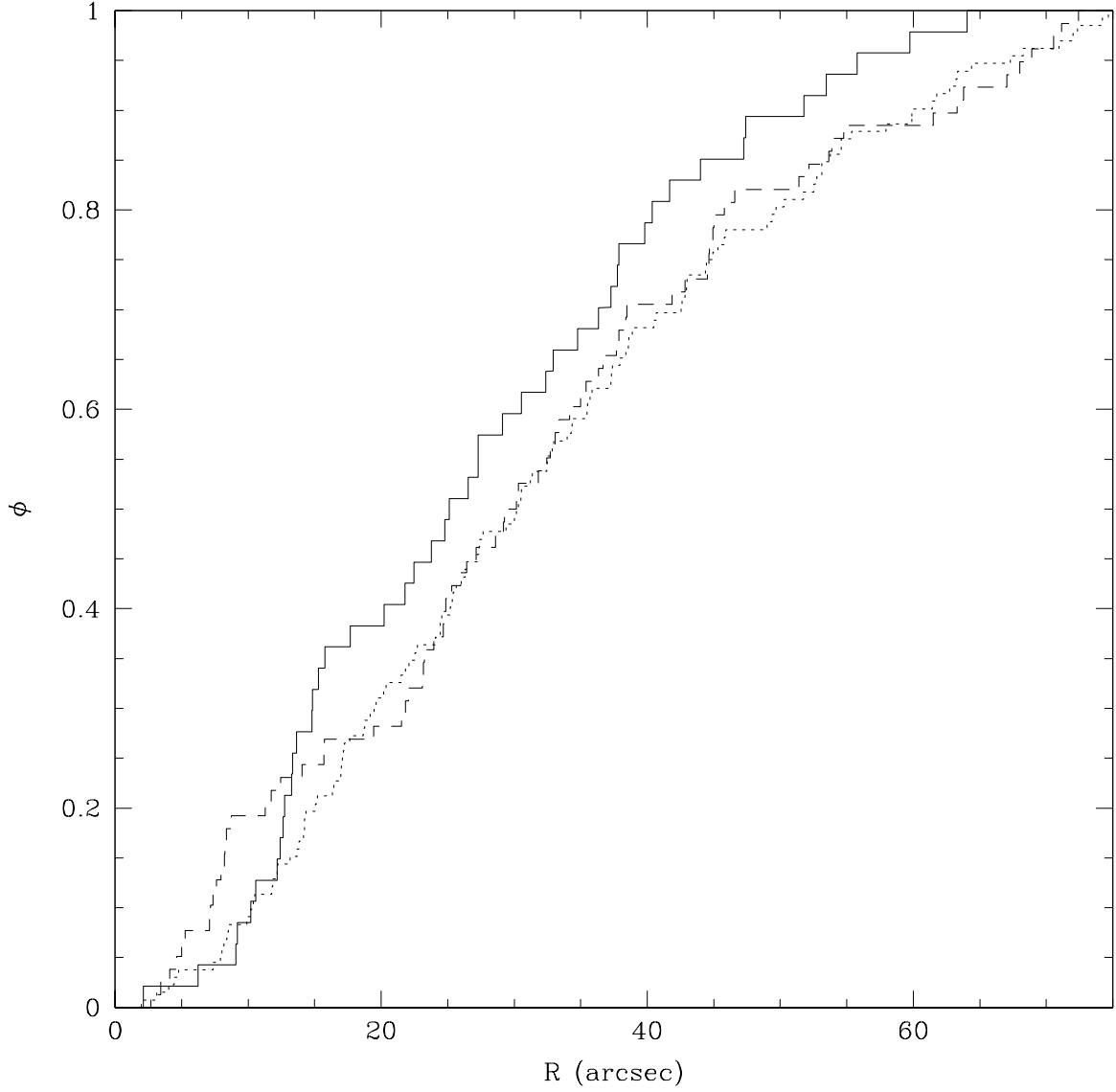}
\caption{Normalized cumulative radial distributions for giant and
subgiant stars ({\it solid line}), probable single main sequence stars
({\it dashed line}), and probable main sequence binary systems ({\it dotted
line}).\label{ccrd}}
\end{figure}

We checked to see whether the identified binary population ($R_{VI} >
0.3$) followed the same radial distribution as the rest of the main
sequence stars.  Because our dataset was the only one appropriate to
carry out this analysis, the cumulative radial distributions shown in
Fig. \ref{ccrd} completely cover only the innermost $75\arcsec$ of the
cluster.  A K-S test indicates that there is a 55\% chance that the
two samples come from the same distribution. There is a similar chance
(51\%) that the subgiant and red giant sample comes from the same
distribution as the combined single-star and binary sequences.  This
analysis is limited by the radial coverage of the cluster in our
images and small-number statistics. However, \citet{koch} found in
their study of Pal 5 that the total binary and single star populations
had similar radial distributions, but brighter photometric binaries
appeared to be more centrally concentrated than fainter ones. If this
also holds for Pal 13, it would help to hide differences between the
binary and single star populations.  Deeper and wider-field images
will be needed to determine whether the binary star radial
distribution deviates from that of the single stars in any substantial way.

\section{Conclusions}\label{5:con}

In this paper we have presented high-precision photometry for stars in
Palomar 13 between the horizontal branch and over 3 magnitudes below
the main sequence turnoff. The photometry has allowed us to
simultaneously compile a complete sample of blue stragglers and
characterize the overall binary fraction for this low-mass globular
cluster. We find that the specific frequency of blue stragglers in
this cluster is one of the highest known. The cumulative
radial distributions of stragglers, subgiant and giant stars, and main
sequence stars each show highly significant differences. The cluster
binary fraction is quite high ($f_b \gtrsim 0.30\pm0.04$), similar to
another low-mass cluster (E3).

Palomar 13 is now one of the few clusters in which the blue straggler
population has been almost completely surveyed {\it and} the binary
fraction has been constrained.  The high binary fraction (probably the
result of mass segregation and tidal evaporation) could be directly
involved in creating the high specific frequency of blue stragglers
seen in the cluster. Davies, Piotto, \& De Angeli (2004) present a
hypothesize that blue straggler populations are produced in collisions
(a mechanism that dominates if $M_V \lesssim -8.8$) or primordial
binary systems. In their scenario, none of Palomar 13's stragglers
would have resulted from collisions.  Palomar 13 would also have
started with a larger binary fraction because a larger proportion of
its binary systems would have avoided breakup in interactions with
other stars.  Davies et al. calculate that the binary fraction would
scale linearly with $\log M_{tot}$ if the initial distribution of
binary separations is uniform in $\log a$, where $a$ is the average
separation.

The situation may not be this simple if Palomar 5 is any indication:
\citet{oden} find a binary fraction of only 10\%, which is lower than
those of both Pal 13 and NGC 288.
In addition, the Davies et al. scenario assumes that globular clusters have
essentially evolved in isolation.  The blue straggler
population of the cluster might be expected to show qualitative
differences depending on its dynamical history --- whether it began
life as a very low-mass globular cluster or if it was initially more
massive and was heavily affected by evaporation of its lower mass
stars. Earlier studies \citep{sie01,cot02} present evidence that
evaporation probably played a major role. If true, then the effects
(if any) would show up in the faint population of stragglers because
their lifetimes are a much larger fraction of the lifetime of the
cluster. If primordial binary stars are largely retained by the
cluster during evaporation, then the blue straggler population of the
cluster should be the equivalent of that of a more massive cluster. A
cluster that had low mass from the beginning would have a smaller
straggler population overall (although this would be partially
compensated for by a larger binary fraction).

On the other hand, the horizontal branch population in Palomar 13 is
roughly consistent with that of a cluster of its {\it current}
mass. \citet{pio04} show that the number of horizontal branch stars
per unit absolute visual flux is very nearly constant over 4.5 mag in
$M_V$. If anything, Palomar 13 has fewer HB stars than expected for its
luminosity. In conjunction with the cumulative radial distributions
(discussed in this paper) and the flatness of the main sequence
luminosity function \citep{cot02}, this indicates that {\it
evaporation is affecting single stars throughout the main sequence and
giant branch of this cluster}. These (admittedly simple) arguments
imply that tidal effects on a cluster may constitute a second-order
influence on the size of a cluster blue straggler population.

\acknowledgements

The authors wish to thank P. C\^ot\'e and M. Siegel for the generous
use of their datasets on the cluster, and the anonymous referee for 
comments that helped improve the manuscript. L.L.C. would like to thank
L. M. Hicks, L. Layton, and N. Fekadu. E.L.S. also wishes to thank
F. De Angeli for helpful conversations. This research was supported by
the National Science Foundation under grant AST 00-98696.

\clearpage


\clearpage

\clearpage

\clearpage

\clearpage

\clearpage

\clearpage

\clearpage

\clearpage

\clearpage

\clearpage

\begin{deluxetable}{rrrrrllcccl}
\tablewidth{0pt}
\tabletypesize{\scriptsize}
\tablecaption{Straggler Star Candidates}
\tablehead{\colhead{ID} & \colhead{SMCT} & \colhead{CDMCM\tablenotemark{1}} & \colhead{$\Delta \alpha$ (\arcsec)} & \colhead{$\Delta \delta$ (\arcsec)} &
\colhead{$V$} & \colhead{$(V-I)$} & \colhead{$(B-V)$} & \colhead{$(U-V)$} & \colhead{$P$} & \colhead{Notes}} 
\startdata
2\tablenotemark{2}   & 79 & 29 &  $-7.38$ & $-28.60$ & $18.939 \pm 0.010$ & $0.526 \pm 0.016$ &       &       &      & rv member\\
                     &    &    &          &          & 18.956             & 0.521             & 0.373 &       &      & \\
                     &    &    &          &          & 19.021             &                   & 0.365 & 0.397 & 0.65 & BSS 9 (SMCT) \\
7\tablenotemark{2}   &125 & 38 & $-11.05$ & $-30.67$ & $19.575 \pm 0.011$ & $0.595 \pm 0.015$ &       &       &      & \\
                     &    &    &          &          & 19.580             & 0.594             & 0.463 &       &      &\\
                     &    &    &          &          & 19.603             &                   & 0.455 & 0.463 & 0.58 & BSS 8 (SMCT) \\
9\tablenotemark{23}  &126 & 40 &   21.70  & $-27.35$ & $19.649 \pm 0.011$ & $0.340 \pm 0.016$ &       &       &      & \\
                     &    &    &          &          & 19.700             & 0.403             & 0.278 &       &      & \\
                     &    &    &          &          & 19.714             &                   & 0.268 & 0.341 & 0.95 & BSS 6 (SMCT) \\
11\tablenotemark{23} &124 & 41 &  $-0.52$ &    1.07  & $19.707 \pm 0.011$ & $0.104 \pm 0.016$ &       &       &      & \\
                     &    &    &          &          & 19.699             & 0.100             & 0.161 &       &      & \\
                     &    &    &          &          & 19.687             &                   & 0.190 & 0.322 & 0.80 & BSS 5 (SMCT) \\
                     &138 & 44 & 75.87    &$-169.75$ & 19.712             & 0.461             & 0.387 &       &      & outside field\\
                     &    &    &          &          & 19.741             &                   & 0.389 & 0.461 & 0.79 & BSS 7 (SMCT) \\
16\tablenotemark{23} &167 & 55 &  $-9.42$ & 12.08    & $19.972 \pm 0.011$ & $0.223 \pm 0.017$ &       &       &      & \\
                     &    &    &          &          & 19.968             & 0.223             & 0.214 &       &      & \\
                     &    &    &          &          & 20.001             &                   & 0.210 & 0.294 & 0.75 & BSS 2 (SMCT) \\
17\tablenotemark{23} &179 & 52 & $-20.09$ & 28.56    & $19.977 \pm 0.011$ & $0.373 \pm 0.017$ &       &       &      & \\
                     &    &    &          &          & 19.932             & 0.346             & 0.295 &       &      & \\
                     &    &    &          &          & 19.991             &                   & 0.264 & 0.329 & 0.81 & BSS 1 (SMCT) \\
29                   &239 & 79 & $-26.38$ & $-0.35$  & $20.471 \pm 0.006$ & $0.623 \pm 0.014$ &       &       &      & binary or BSS?\\
                     &    &    &          &          & 20.475             & 0.612             & 0.520 &       &      & \\
                     &    &    &          &          & 20.374             &                   & 0.427 & 0.347 & 0.97 &  \\
36                   &289 & 99 & $-27.13$ & $-74.02$ & $20.660 \pm 0.005$ & $0.370 \pm 0.015$ &       &       &      & \\
                     &    &    &          &          & 20.650             & 0.394             & 0.337 &       &      & \\
                     &    &    &          &          & 20.680             &                   & 0.331 & 0.252 & 0.98 &  \\
40\tablenotemark{23} &306 & 98 & $-6.25$  & 15.36    & $20.715 \pm 0.006$ & $0.548 \pm 0.015$ &       &       &      & \\
                     &    &    &          &          & 20.720             & 0.534             & 0.426 &       &      & \\
                     &    &    &          &          & 20.760             &                   & 0.397 & 0.069 & 0.87 & BSS 3 (SMCT) \\
42                   &356 &100 & 31.22    & $-13.09$ & $20.731 \pm 0.006$ & $0.426 \pm 0.016$ &       &       &      & \\
                     &    &    &          &          & 20.744             & 0.458             & 0.342 &       &      &  \\
                     &    &    &          &          & 20.851             &                   & 0.373 & 0.269 & 0.91 &  \\
43                   &352 &108 & 2.87     & $-27.79$ & $20.749 \pm 0.006$ & $0.471 \pm 0.015$ &       &       &      & member?\\
                     &    &    &          &          & 20.746             & 0.442             & 0.374 &       &      & \\
                     &    &    &          &          & 20.885             &                   & 0.326 & 0.154 & 0.49 &  \\
45                   &366 &110 & 13.09    & $-22.18$ & $20.747 \pm 0.006$ & $0.539 \pm 0.015$ &       &       &      & member?\\
                     &    &    &          &          & 20.755             & 0.549             & 0.417 &       &      & \\
                     &    &    &          &          & 20.793             &                   & 0.460 & 0.418 & 0.46 &  \\
52                   &    &118 & 3.28     & $-12.00$ & $20.822 \pm 0.006$ & $0.504 \pm 0.015$ &       &       &      & unresolved blend in SMCT\\
                     &    &    &          &          & 20.817             & 0.469             & 0.382 &       &      & \\
53                   &364 &123 & $-20.51$ & 5.12     & $20.845 \pm 0.006$ & $0.558 \pm 0.015$ &       &       &      & \\
                     &    &    &          &          & 20.843             & 0.563             & 0.457 &       &      & \\
                     &    &    &          &          & 20.827             &                   & 0.384 & 0.480 & 0.57 &  \\
64                   &497 &148 & $-35.63$ & 19.66    & $21.058 \pm 0.006$ & $0.507 \pm 0.018$ &       &       &      & \\
                     &    &    &          &          & 21.057             & 0.520             & 0.396 &       &      & \\
                     &    &    &          &          & 21.058             &                   & 0.434 & 0.553 & 0.86 &   \\
67                   &504 &158 & $-31.85$ & $-18.55$ & $21.078 \pm 0.006$ & $0.525 \pm 0.016$ &       &       &      & field?\\
                     &    &    &          &          & 21.090             & 0.507             & 0.433 &       &      & \\
                     &    &    &          &          & 21.104             &                   & 0.371 & 0.207 & 0.07 &  \\
75                   &524 &167 & $-10.24$ & $-11.64$ & $21.210 \pm 0.006$ & $0.519 \pm 0.019$ &       &       &      & \\
                     &    &    &          &          & 21.218             & 0.499             & 0.402 &       &      & \\
                     &    &    &          &          & 21.132             &                   & 0.468 & 0.339 & ...  &   \\
77                   &623 &180 &$-25.92$  & 9.65     & $21.253 \pm 0.006$ & $0.589 \pm 0.009$ &       &       &      & \\
                     &    &    &          &          & 21.262             & 0.642             & 0.486 &       &      & turnoff star?\\
                     &    &    &          &          & 21.249             &                   & 0.515 & 0.577 & 0.59 & \\
78                   &598 &172 & $-6.49$  & 2.99     & $21.272 \pm 0.006$ & $0.524 \pm 0.009$ &       &       &      & \\
                     &    &    &          &          & 21.279             & 0.546             & 0.436 &       &      & \\
                     &    &    &          &          & 21.310             &                   & 0.476 & 0.099 & 0.85 &   \\
80                   &693 &183 & $-12.35$ & $-12.62$ & $21.315 \pm 0.006$ & $0.565 \pm 0.016$ &       &       &      & \\
                     &    &    &          &          & 21.317             & 0.560             & 0.464 &       &      & \\
                     &    &    &          &          & 21.389             &                   & 0.506 & 0.239 & ...  &   \\
82                   &721 &182 & 15.82    & $-7.48$  & $21.324 \pm 0.006$ & $0.563 \pm 0.009$ &       &       &      & member?\\
                     &    &    &          &          & 21.319             & 0.538             & 0.441 &       &      & \\
                     &    &    &          &          & 21.362             &                   & 0.596 & 0.249 & 0.51 &  \\
84                   &461 &190 & 2.37     & $-3.17$  & $21.367 \pm 0.006$ & $0.571 \pm 0.009$ &       &       &      &  \\
                     &    &    &          &          & 21.358             & 0.570             & 0.433 &       &      & \\
                     &    &    &          &          & 21.080             &                   & 0.433 & 0.148 & ...  &   \\
112                  &1041&255 & 26.59    & $-80.51$ & $21.776 \pm 0.007$ & $0.631 \pm 0.009$ &       &       &      &  \\
                     &    &    &          &          & 21.768             & 0.689             & 0.484 &       &      & \\
                     &    &    &          &          & 21.773             &                   & 0.463 & 0.215 & 0.53  &   \\
116                  &    &275 & $-8.74$  & 19.62    & $21.867 \pm 0.007$ & $0.636 \pm 0.010$ &       &       &      &  \\
                     &    &    &          &          & 21.872             & 0.679             & 0.523 &       &      & \\
117                  &1169&279 & $-13.25$ & $-7.20$  & $21.877 \pm 0.007$ & $0.610 \pm 0.010$ &       &       &      &  \\
                     &    &    &          &          & 21.871             & 0.659             & 0.522 &       &      & \\
                     &    &    &          &          & 22.043             &                   & 0.358 &$-0.134$& 0.17  &   \\
\multicolumn{10}{c}{Stars Redder than Turnoff} \\ 
18                   &202 & 58 & 33.56    & 9.54     & $20.041 \pm 0.011$ & $0.711 \pm 0.016$ &       &       &      & binary?\\
                     &    &    &          &          & 20.048             & 0.709             & 0.520 &       &      & \\
                     &    &    &          &          & 20.038             &                   & 0.620 & 0.547 & 0.85 &  \\
                     &214 & 61 & $-24.19$ & $-87.54$ & 20.060             & 0.776             & 0.590 &       &      & binary?; outside field\\
                     &    &    &          &          & 20.121             &                   & 0.590 & 0.497 & 0.99 &  \\
24                   &    & 71 & $-13.88$ & 1.08     & $20.309 \pm 0.006$ & $0.772 \pm 0.014$ &       &       &      & unresolved blend in SMCT\\
                     &    &    &          &          & 20.300             & 0.757             & 0.593 &       &      & \\
25                   &218 & 70 & $-10.67$ & 15.13    & $20.317 \pm 0.005$ & $0.711 \pm 0.013$ &       &       &      & prob. binary\\
                     &    &    &          &          & 20.317             & 0.702             & 0.529 &       &      & \\
                     &    &    &          &          & 20.243             &                   & 0.504 & 0.301 & 0.94 &  \\
26                   &274 & 75 & 26.88    & $-6.22$  & $20.419 \pm 0.006$ & $0.671 \pm 0.014$ &       &       &      & prob. binary\\
                     &    &    &          &          & 20.425             & 0.678             & 0.532 &       &      & \\
                     &    &    &          &          & 20.413             &                   & 0.604 & 0.541 & 0.91 &  \\
33                   &301 & 86 & $-14.45$ & $-3.34$  & $20.508 \pm 0.006$ & $0.697 \pm 0.014$ &       &       &      & prob. binary\\
                     &    &    &          &          & 20.512             & 0.671             & 0.532 &       &      & \\
                     &    &    &          &          & 20.608             &                   & 0.434 & 0.310 & 0.90 &  \\
\multicolumn{10}{c}{Probable Red Giant Binary Stars} \\ 
3                    & 88 & 30 & $-14.44$ & $-41.58$ & $18.966 \pm 0.010$ & $0.971 \pm 0.015$ &       &       &      & \\
                     &    &    &          &          & 18.976             & 0.970             & 0.749 &       &      & \\
                     &    &    &          &          & 18.970             &                   & 0.757 & 0.872 & 0.93 & \\
14                   &182 & 49 & 3.06     & $-21.59$ & $19.751 \pm 0.011$ & $0.923 \pm 0.016$ &       &       &      & \\
                     &    &    &          &          & 19.758             & 0.917             & 0.716 &       &      & \\
                     &    &    &          &          & 19.795             &                   & 0.644 & 0.592 & 0.98 & \\
\multicolumn{10}{c}{High-Probability Field Stars} \\
                     & 76 & 26 & $-12.51$ & $-92.52$ & 18.758             & 0.877             & 0.744 &       &      & rv nonmember; outside field\\
                     &    &    &          &          & 18.743             &                   & 0.777 & 1.018 & 0.00 &  \\
                     &154 & 45 & 68.75    &$-116.52$ & 19.692             & 0.753             & 0.547 &       &      & outside field\\
                     &    &    &          &          & 19.715             &                   & 0.584 & 0.529 & 0.00 &  \\
                     &168 & 43 &$-178.63$ &$-134.63$ & 19.659             & 0.859             & 0.706 &       &      & outside field\\
                     &    &    &          &          & 19.704             &                   & 0.594 & 0.511 & 0.00 &  \\
\enddata
\label{tab5}
\tablenotetext{1}{\citet{cot02}.}
\tablenotetext{2}{Blue straggler star candidate from Siegel et
al. (2001; SMCT).}  
\tablenotetext{3}{Blue straggler star candidate from Borissova et
al.(1997; BMS).}
\end{deluxetable}

\begin{deluxetable}{cccc}
\tablewidth{0pt} 
\tablecaption{K-S Test Results for Radial Distribution Comparisons} 
\tablehead{\colhead{Sample 1} & \colhead{Sample 2} & \colhead{$D$} &
\colhead{$P$}} 
\startdata
\multicolumn{4}{c}{\citet{cot02} Samples}\\
BSS & GB & 0.42 & 0.00042 \\
BSS & MS & 0.62 & $5 \times 10^{-11}$ \\
GB & MS & 0.23 & 0.00060 \\
BSS $\left[(V-I) < (V-I)_{TO}\right]$ & GB & 0.42 & 0.0011 \\
BSS $\left[(V-I) < (V-I)_{TO}\right]$ & MS & 0.63 & $2 \times 10^{-9}$ \\
BSS $\left[V > 20.4\right]$ & GB & 0.43 & 0.0041 \\
BSS $\left[V > 20.4\right]$ & MS & 0.64 & $2 \times 10^{-7}$ \\
BSS $\left[V < 20\right]$ & GB & 0.44 & 0.12 \\
BSS $\left[V < 20\right]$ & MS & 0.63 & 0.0036 \\
\multicolumn{4}{c}{ESI Field Samples}\\
Binaries & MS & 0.11 & 0.55 \\
RGB & MS+Binaries & 0.13 & 0.51 \\
\enddata
\label{tabks}
\end{deluxetable}

\begin{deluxetable}{rrrrrrrrr}
\tablewidth{0pt}
\tablecaption{Keck ESI Photometry of Palomar 13}
\tablehead{\colhead{ID} & \colhead{$\Delta \alpha (\arcsec$)} & 
\colhead{$\Delta \delta (\arcsec$)} & \colhead{$X$ (pix)} & 
\colhead{$Y$ (pix)} & \colhead{$V$} & \colhead{$\sigma_V$} &
\colhead{$I$} & \colhead{$\sigma_I$}}
\startdata
     1 & $  -2.09$ & $  12.00$ &  284.92 &  514.71 &  18.805 &   0.010 &  17.784 &   0.011\\
     2 & $  -7.38$ & $ -28.61$ &  286.72 &  780.19 &  18.939 &   0.010 &  18.413 &   0.012\\
     3 & $ -14.43$ & $ -41.58$ &  321.84 &  868.99 &  18.966 &   0.010 &  17.996 &   0.011\\
     4 & $  -5.35$ & $  -7.49$ &  290.46 &  642.66 &  19.017 &   0.011 &  18.023 &   0.011\\
     5 & $   0.86$ & $  12.71$ &  266.54 &  507.95 &  19.060 &   0.011 &  18.052 &   0.011\\
     6 & $ -32.40$ & $  40.37$ &  502.77 &  354.47 &  19.355 &   0.010 &  18.382 &   0.011\\
     7 & $ -11.05$ & $ -30.67$ &  308.69 &  796.20 &  19.575 &   0.011 &  18.980 &   0.011\\
     8 & $ -12.17$ & $   2.39$ &  342.20 &  584.09 &  19.598 &   0.011 &  18.635 &   0.012\\
     9 & $  21.70$ & $ -27.35$ &  100.45 &  750.53 &  19.649 &   0.011 &  19.309 &   0.012\\
    10 & $  11.46$ & $ -38.14$ &  157.82 &  827.63 &  19.636 &   0.011 &  18.677 &   0.011\\
    11 & $  -0.52$ & $   1.07$ &  266.15 &  583.94 &  19.707 &   0.011 &  19.602 &   0.012\\
    12 & $ -20.27$ & $  -9.65$ &  384.82 &  667.65 &  19.703 &   0.011 &  18.747 &   0.012\\
    13 & $  12.57$ & $   4.24$ &  184.34 &  553.86 &  19.719 &   0.011 &  18.766 &   0.011\\
    14 & $   3.06$ & $ -21.59$ &  225.04 &  727.25 &  19.750 &   0.011 &  18.828 &   0.012\\
\enddata
\label{tab6}
\tablecomments{The complete version of this table is in the
electronic edition of the Journal. The printed edition contains only a sample.}
\end{deluxetable}

\end{document}